# Quantitative First-Pass Perfusion CMR: from technical principles to clinical practice

Catarina N Carvalho [a,b], Andreia Gaspar* [b], Carlos Real* [c,d], Carlos Galán-Arriola* [c,e], Elisa Moya-Sáez* [f], Rosa-María Menchón-Lara [f,g], Javier Sanchez [h], Carlos Alberola-López [f], Rita G Nunes [b], Borja Ibáñez [c,e,i], Teresa M Correia [b,j]

[a] Centre of Marine Sciences (CCMAR), Universidade do Algarve, Campus de Gambelas, Faro, Portugal; [b] Institute for Systems and Robotics - Lisboa and Department of Bioengineering, Instituto Superior Técnico – Universidade de Lisboa, Lisbon, Portugal; [c] Centro Nacional de Investigaciones Cardiovasculares (CNIC), Madrid, Spain; [d] Department of Cardiology, Hospital Universitario Clínico San Carlos, Madrid, Spain; [e] Centro de Investigación Biomédica en Red en Enfermedades Cardiovasculares (CIBERCV), Madrid, Spain; [f] ETSI Telecomunicación. Universidad de Valladolid, Valladolid, Spain; [g] Universidad Politécnica de Cartagena, Murcia, Spain; [h] Philips, Madrid, Spain; [i] Cardiology Department, IIS-Fundación Jiménez Díaz Hospital, Madrid, Spain; [j] School of Biomedical Engineering and Imaging Sciences, King's College London, London, United Kingdom

* equal contribution

14 February 2025

**Abstract**

Myocardial perfusion cardiovascular magnetic resonance (pCMR) using first-pass contrast-enhanced imaging could play an important role in the detection of epicardial and microvascular coronary artery disease. Recently, the emergence of quantitative pCMR has provided a more reliable and observer-independent analysis compared to visual interpretation of dynamic images. This review aims to cover the basics of quantitative pCMR, from acquisition protocols, its use in preclinical and clinical studies, image reconstruction and motion handling, to automated quantitative pCMR pipelines. It also offers an overview of emerging tools in the field, including artificial intelligence-based methods.

**Acronyms**

ACS: Acute coronary syndrome

ADMM: Alternating direction method of multipliers

AHA: American Heart Association

AI: Artificial intelligence

AIF: Arterial input function

bSSFP: Balanced steady state free precession

BTEX: Blood tissue exchange

CAD: Coronary artery disease

CCS: Chronic coronary syndrome

CMD: Coronary microvascular dysfunction

CMR: Cardiovascular magnetic resonance

CMRA: Coronary Magnetic Resonance Angiography

CNN: Convolutional neural network

CS: Compressed sensing

CT: Computed tomography

DCE: Dynamic contrast enhanced

DL: Deep learning

ECG: Electrocardiogram

FISTA: Fast iterative shrinkage/thresholding algorithm

GRE: Gradient echo

MBF: Myocardial blood flow

MINOCA: Myocardial infarction with non-obstructive coronary arteries

MR: Magnetic resonance

PCA: Principal component analysis

pCMR: Perfusion cardiovascular magnetic resonance

PD: Proton density

PET: Positron emission tomography

PI: Parallel imaging

RF: Radiofrequency

RNN: Recurrent neural network

SMS: Simultaneous multi-slice

SNR: Signal-to-noise ratio

SPECT: Single-photon emission computed tomography

SSDU: Self-supervised learning via data undersampling

TE: Echo time

TS: Saturation time

# 1. Introduction

First-pass perfusion cardiovascular magnetic resonance (pCMR) is an established non-invasive and radiation-free imaging technique that allows the detection of reduced blood flow to the heart, i.e. myocardial ischaemia, with superior accuracy compared to single-photon emission computed tomography (SPECT), and comparable accuracy to positron emission tomography (PET) and invasive intracoronary methods (1–6). Current clinical practice typically involves the visual assessment of a series of contrast-enhanced dynamic images by well-trained and experienced reporters, which is observer- and scanner-dependent and does not provide quantifiable metrics, limiting the widespread adoption of pCMR. More recently, quantitative pCMR has emerged as a more reliable and operator-independent approach for identifying perfusion defects.

This review aims to provide an overview of quantitative pCMR and to highlight key emerging techniques in the field. **Figure 1** showcases the state-of-the-art steps for a quantitative pCMR exam and the areas where improvements can be made. Each step is described in a section of this review. Section 2 describes the clinical motivation behind using CMR. Section 3 describes the guidelines for preclinical and clinical studies in pCMR. Section 4 provides an overview of the data acquisition sequences for quantitative pCMR. Section 5 overviews the tracer-kinetic models commonly employed to obtain quantitative perfusion measurements. Section 6 explains the need for acceleration in pCMR and provides an overview of the techniques developed for this purpose, and motion-correction tools that may be employed. Section 7 briefly introduces the main principles and concepts behind artificial intelligence and reviews architectures developed for varied purposes in pCMR. Finally, Section 8 describes different approaches proposed for automating the full pCMR pipeline. Some of these topics have been individually covered in detail in other review papers, for example, in (7–13), but a comprehensive review of the entire pipeline is still lacking.

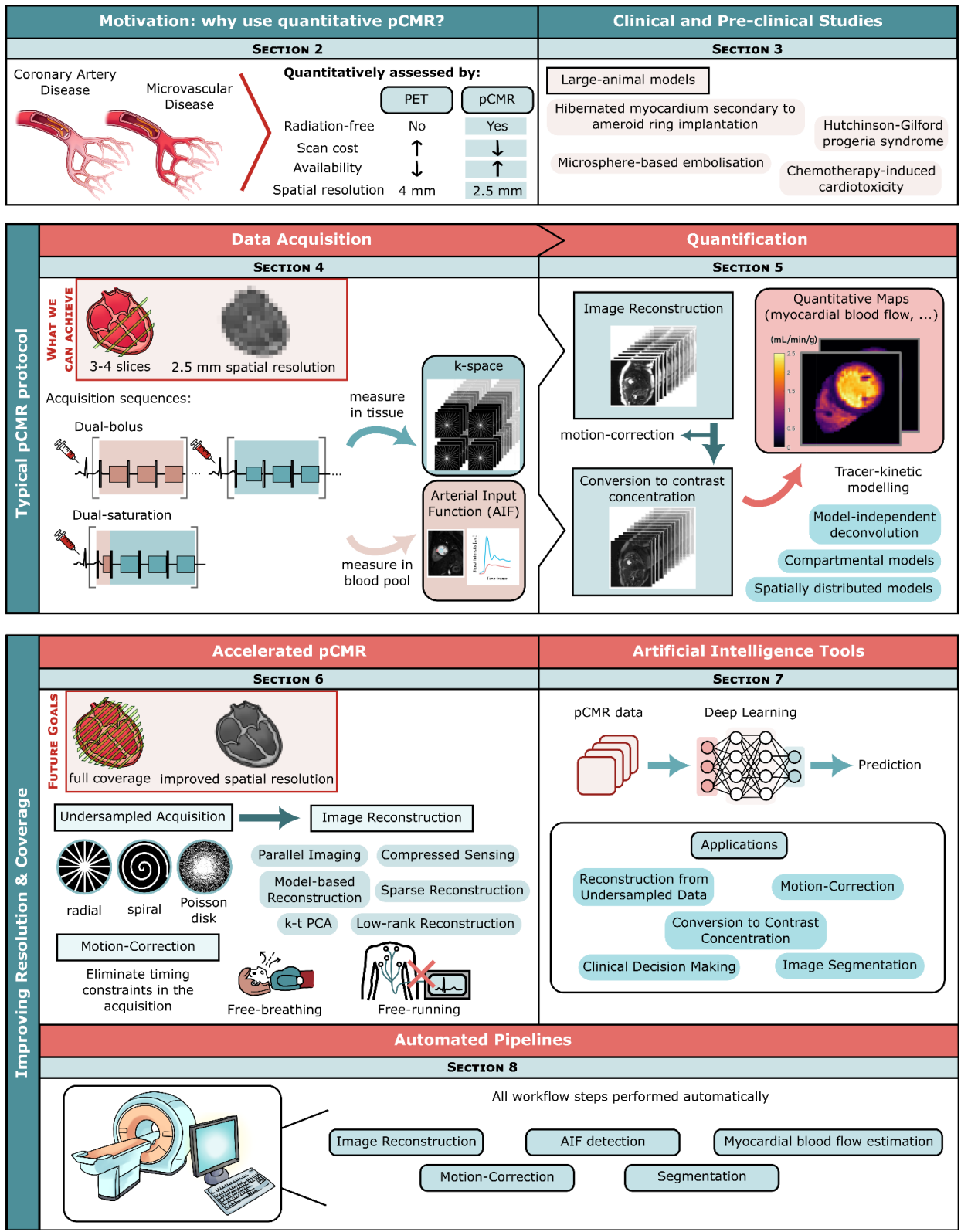

**Figure 1.** Overview of a quantitative pCMR exam. (2) Quantitative pCMR can evaluate coronary artery and microvascular diseases, with advantages over PET regarding the use of

Quantitative First-Pass Perfusion CMR, Feb 2025

ionizing radiation, scan cost, availability and image resolution. A typical pCMR protocol is electrocardiogram-synchronised, requires a long breath-hold, achieves a heart coverage of 3-4 slices and in-plane spatial resolution of approximately 2.5 mm, and involves (4) a data acquisition step that obtains both the k-space data and the arterial input function, typically either with a dual-bolus or a dual-saturation acquisition sequence; (5) a quantification step, where the acquired data are first reconstructed into a series of contrast-weighted images, then converted into contrast agent concentration, and finally into quantitative maps, by application of a tracer-kinetic model. (6) The main desired improvements on the standard pCMR protocol consist in increasing the heart coverage and improving the spatial resolution. This can be achieved by accelerating the acquisition, through k-space undersampling and subsequent reconstruction with strategies such as parallel imaging and compressed sensing. Acquisitions can also be performed under free-breathing and free-running conditions (electrocardiogram-free), by applying motion-correction strategies. (7) Additionally, artificial intelligence can be employed to accelerate both the data acquisition and the quantification steps. (8) Finally, there is growing interest in the implementation of automated pipelines that perform image reconstruction, motion correction, image segmentation, arterial input function detection, and myocardial blood flow estimation in-line on the scanner or dedicated software platforms for pCMR analysis.

# 2. Why perfusion CMR? Clinical motivation and guidelines

Cardiac imaging has become an indispensable diagnostic tool. Risk stratification and treatment options rely to a great extent on the results of cardiac imaging exams. Blood supply to the myocardium (perfusion) is affected in several cardiac diseases (14); it can be affected in basal conditions (i.e. present at rest) or it can become apparent when there is an increase in myocardial blood demand (i.e. during stress). Perfusion abnormalities can be the result of defects at different levels, ranging from epicardial coronary artery stenoses to microcirculation defects, or the combination of both. Coronary macro- (epicardial) and micro-circulation defects can be structural (e.g. destruction of capillary) or functional (e.g. lack of dilation capacities) (14).

There are several techniques available to detect myocardial perfusion defects. Myocardial perfusion can be assessed invasively (i.e. cardiac catheterisation with insertion of intracoronary pressure/flow wires) or non-invasively by using PET, SPECT, computed tomography (CT), and cardiac magnetic resonance (CMR). Some of these techniques allow only a visual assessment of perfusion, usually using stress imaging. However, there is increasing evidence that perfusion evaluation should be based on quantitative analysis rather than visual assessment to detect a perfusion defect. PET allows non-invasive quantitative assessment of myocardial perfusion, since myocardial blood flow can be measured in absolute units (ml/min/g). This quantitative parameter allows identifying reduced myocardial flow even in complex situations such as coronary artery stenosis in different vessels or in those with microvascular dysfunction (15). These advantages have made the diagnostic performance of PET superior to visual assessment with other imaging techniques such as SPECT (2). However, the short half-life of commonly used tracers makes myocardial perfusion PET impractical and has hampered its widespread use.

Among all imaging modalities, CMR provides one of the most comprehensive exams for cardiovascular assessment, as it can accurately evaluate anatomy, function, viability, tissue composition and blood flow (16), in a single radiation-free imaging session. CMR is known to be the gold standard imaging method to assess ventricular size and function, as well as myocardial viability in patients with coronary artery disease (CAD) (17). The higher spatial and temporal resolution of CMR compared to other imaging techniques makes it the ideal test, for instance, to assess wall motion abnormalities in response to stress. Finally, in recent years, myocardial pCMR based on visual assessment has shown to have better accuracy for diagnosis and exclusion of significant CAD compared to SPECT imaging (18). Currently, there has been a growing interest in developing automated pCMR methods that quantify myocardial perfusion (19,20) (**Figure 2**). In fact, sex- and age- reference values for automated quantitative pCMR were recently published (21). Considering that absolute values of myocardial blood flow can be obtained, both at rest and under stress, the same advantages previously mentioned for PET are anticipated for pCMR. Thus, the continuous evolution of this technology is making it the ideal imaging test to evaluate patients with suspected perfusion defects in a comprehensive manner (**Figure 3**).

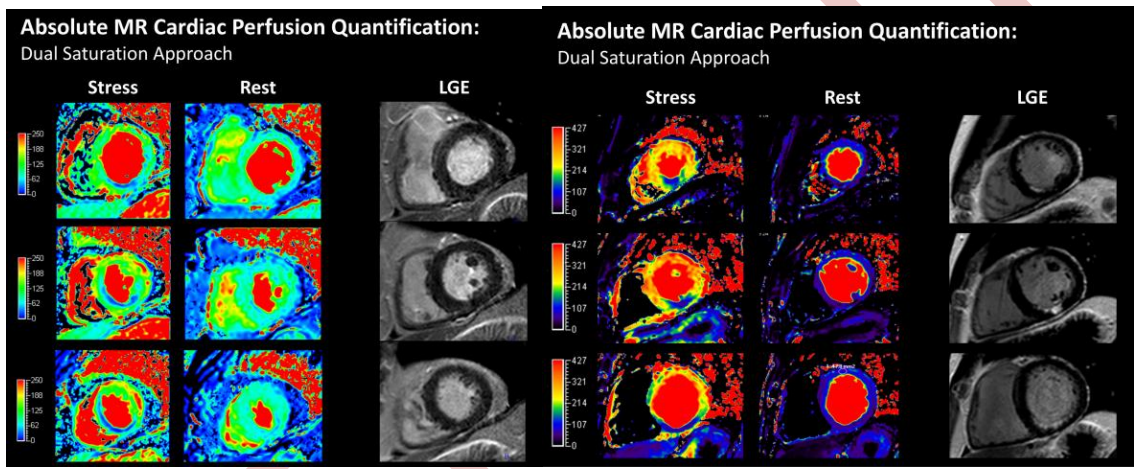

**Figure 2. Examples of pCMR myocardial maps in mL/min/100g.** (left) Patient with endocardial scar tissue in the late gadolinium enhancement (LGE) images in the inferior region and with a clear transmural perfusion defect in stress perfusion. (right) Patient without scar tissue in the LGE images showing an endocardial perfusion defect in the inferior and inferioseptal region in the stress images, but not in the rest pCMR, indicating the presence of inducible ischaemia.

Epicardial CAD is the most important cause of perfusion impairment because of its magnitude, prognostic impact and therapeutic options (22). According to both European and American clinical practice guidelines, for patients with intermediate-risk in whom CAD cannot be excluded by clinical assessment, non-invasive diagnostic tests are recommended. On the other hand, for patients with high clinical likelihood of CAD or an acute coronary syndrome (ACS), proceeding directly to invasive coronary angiography is recommended (22,23). However, there are patients who present with ACS and even with myocardial infarction despite having non-obstructive coronary arteries. The latter situation, known as myocardial infarction with non-obstructive coronary arteries (MINOCA), is relatively frequent (up to 14% of patients with ACS undergoing coronary angiography) and represents a diagnostic challenge (24). In this context, CMR is one of the key diagnostic tools to determine the underlying cause, being the gold standard for the diagnosis of diseases that can present as MINOCA, such as myocarditis (25). Furthermore, quantitative myocardial perfusion diagnostic methods can detect a globally decreased coronary flow reserve, which may indicate coronary microvascular dysfunction (CMD) (15). The latter is a cause of MINOCA and is known to have prognostic implications, thus being an important diagnosis to consider (26). Also, structural alterations in some cardiac diseases such as hypertrophic cardiomyopathy, diabetic and hypertensive heart disease are known to cause CMD (14,27). Finally, CMD was independently associated with cardiovascular and heart failure-related events at follow-up in patients with heart failure and preserved left ventricular ejection fraction (28). Therefore, the availability of a non-invasive quantitative diagnostic method that reliably assesses myocardial perfusion is of paramount importance.

In summary, the development of accurate CMR methodologies for quantitative myocardial perfusion assessment is a clear clinical need that would help clinicians in the management of patients with several conditions, in particular those with suspected CAD. It would be ideal to automate and standardise quantitative pCMR to enable accurate and reproducible measurements of myocardial blood flow and thus, facilitate its adoption in routine clinical practice.

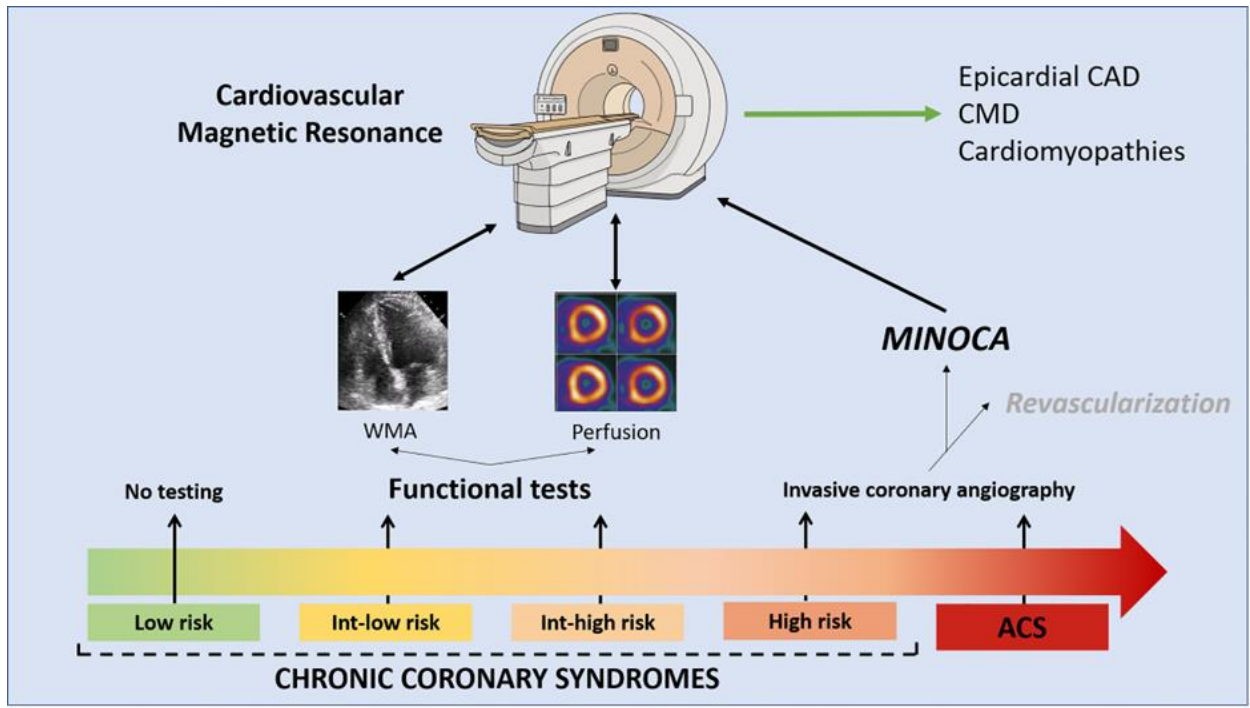

**Figure 3. Usefulness of CMR in patients with suspected coronary artery disease.** According to international guidelines, patients with chronic coronary syndrome (CCS) and low suspicion of having ischemic disease do not require further study. However, patients presenting with CCS and intermediate (Int) risk of having CAD may undergo functional tests. Finally, patients with high risk of having CAD or those presenting with ACS should undergo directly invasive coronary angiography. CMR, specially with quantitative perfusion, could cover all this range of patients. ACS: acute coronary syndrome, CAD: coronary artery disease, CMD: coronary microvascular disease, MINOCA: myocardial infarction with non-obstructive coronary arteries, WMA: wall motion abnormalities.

# 3. The use of quantitative pCMR in Preclinical Studies

Large animal models (Figure 4) are ideal for testing diagnostic and therapeutic options in experimental settings as close as possible to humans. Primates, pigs, sheep and dogs have all been used for testing myocardial imaging protocols. With the exception of primates (not widely used for ethical reasons), the pig is the best candidate species due to its similarities in cardiac anatomy and physiology with humans (29). The coronary anatomy of pigs closely resembles that of humans, although they have fewer than 10% of the coronary collaterals typically found in humans. In both, there is prominent left coronary artery circulation, and the coronary arteries arise in a similar manner; the presence and anatomy of septal and interventricular branches from the left anterior descending artery are the major differences based on atrial circulation. Several human cardiac conditions can be modelled in pigs, such as acute myocardial infarction, hibernating myocardium, damage to coronary microcirculation, persistent atrial fibrillation and Timothy syndrome 1 (29–34).

There are several large-animal models of CAD and microvascular dysfunction, including the hibernated myocardium secondary to ameroid ring implantation (33), microsphere-based embolisation (35) and chemotherapy-induced cardiotoxicity (36). In recent years, transgenic disease models involving microvascular impairment, such as diabetic or Hutchinson-Gilford progeria syndrome pig, have been developed (37,38). However, studies evaluated with quantitative pCMR sequences are lacking and most of the existing ones are only conducted for validating the technique in the acute myocardial infarction model.

Quantitative pCMR has emerged as one of the exploratory endpoints in the cardioprotection field after ischemia/reperfusion in both clinical and preclinical trials (39). In the last few years, quantitative pCMR has improved its methodology in the translational scenario (19). Regarding large-animal studies, it has been mostly used to monitor response to stem cell therapy (40,41) as a surrogate parameter for neovascularization after paracrine effect of the injected cells. Quantitative pCMR has also been demonstrated useful for subtle cardiac damage evaluation in different models of disease including microembolisation with normal epicardial coronary arteries (42), Hutchinson-Gilford progeria syndrome (38) or anthracycline-induced cardiotoxicity (36).

Stress quantitative pCMR has been tested in pig studies, specifically in the context of progressive coronary artery obstruction (43), in which Van Houten and colleagues showed early perfusion defects at stressed conditions after ameroid coronary surgery in the circumflex coronary artery. Interestingly, they found similar results regarding hyperemia in pigs as in other studies where higher doses of adenosine, along with drugs to stabilise arterial pressure, were required to achieve proper levels of myocardial hyperemia compared to humans (44).

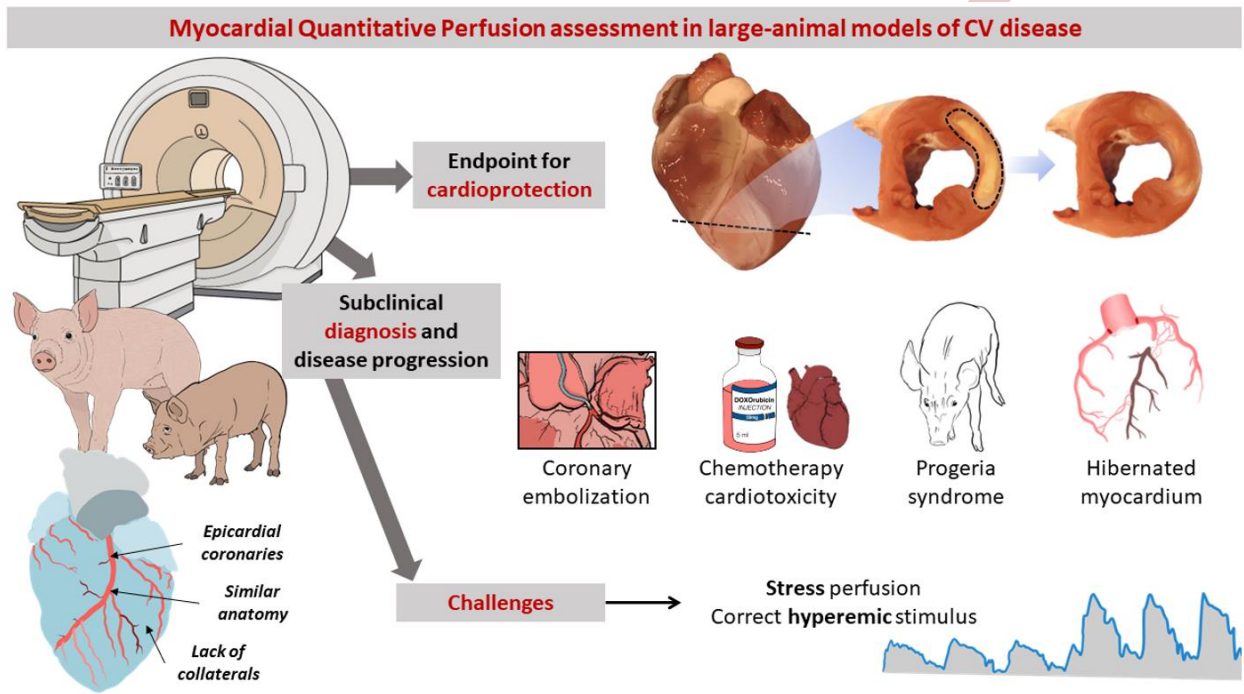

**Figure 4. Large-animal models in myocardial quantitative perfusion.** pCMR is typically used not only for model characterization but also as a secondary/exploratory endpoint in preclinical trials for cardioprotection. Currently, the main challenge is related with the fine tuning of reliable and reproducible stress protocols.

# 4. Sequences for Quantitative pCMR: how to obtain data for accurate quantification of myocardial perfusion

First-pass pCMR is obtained by imaging the heart during the first pass of a contrast agent (e.g. Gadolinium [Gd]) with a lower longitudinal relaxation constant (T1) than the myocardium. When using a magnetic resonance (MR) sequence that is sensitive to T1, the measured signal intensity will relate to the contrast concentration in the different regions of the myocardium (45). The sensitivity to T1 can be enhanced by applying a saturation radiofrequency (RF) pulse and subsequently acquiring the image with a specific saturation delay. This approach provides qualitative information of the contrast dynamics. However, to obtain quantitative pCMR maps, it is necessary to convert the MR signal intensity changes into myocardial blood flow (MFB).

To correctly assess MBF (in ml/min/g), it is necessary to know the arterial input function (AIF), a measure of the contrast agent concentration that is supplied to the heart (46). The AIF can be estimated by measuring the signal intensity curve on the blood, typically on the output track of the aorta (47), since the blood flow influences the MR signal. However, in perfusion measurement conditions (i.e. long saturation time and high contrast concentration in the aorta), the longitudinal magnetization of the blood will be fully recovered (resulting in signal saturation); in addition, it may suffer from T2* related signal loss, leading to underestimation of the AIF (48). To accurately measure the AIF, two solutions have been proposed, which measure it in different conditions: lower contrast dose (e.g. dual-bolus method (49)) or lower saturation time (TS) (e.g. dual-sequence or dual-saturation method (19,50)).

The dual-bolus sequence (49) consists in performing two contrast injections with the same volume but with a different dose: a low-dose injection to obtain a non-fully recovered signal in the blood pool, and a second full-dose injection to obtain conventional information from the myocardium (**Figure 5-a**). An additional benefit of the dual-bolus sequence is the reduction of T2* effects in the AIF estimation due to lower contrast concentration. One disadvantage of this technique is that the AIF and perfusion images are acquired using two different boluses, so there may be differences regarding cardiac heart rate and/or respiratory motion, which can lead to errors in the estimation.

The other approach, the dual-sequence or dual-saturation sequence (19,50), consists in acquiring a low-resolution image with low echo time (TE) (to reduce T2*-related signal loss) and short TS for AIF estimation (**Figure 5-b**). This image can be obtained at the beginning of each cardiac cycle, before the high-resolution T1-weighted perfusion images are acquired. This can also be combined with a dual-echo acquisition to measure the T2* during the first pass of the contrast bolus (51). This solution avoids the injection of an additional contrast bolus, but requires specialised software to perform the acquisition.

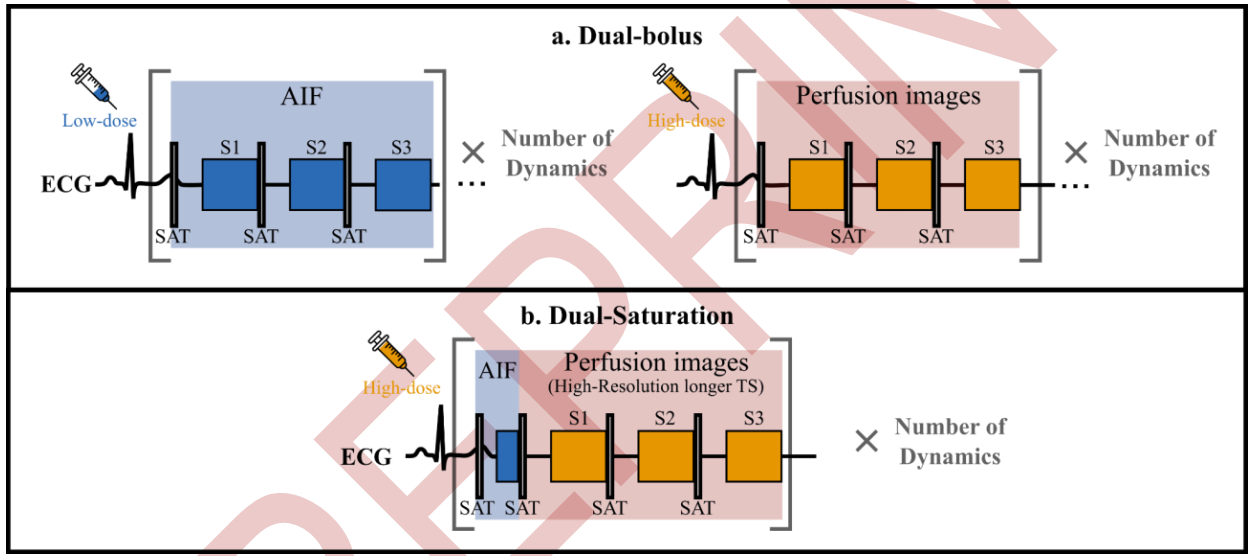

**Figure 5. T1-weighted MR sequences for quantitative first-pass pCMR.** Two methodologies are currently available: **a)** the dual-bolus method first uses a low dose bolus to measure the AIF followed by a high dose bolus for myocardial imaging; **b)** the dual-saturation method includes the acquisition of a low-resolution AIF image with short saturation time, followed by high-resolution myocardial perfusion images acquired with a longer saturation time, within the same cardiac cycle.

Dynamic T1-weighted MR sequences are used in pCMR to capture the passage of the contrast bolus. Balanced steady state free precession (bSSFP) sequences can be used to capture the maximum information needed because this type of readout is fast and provides a high signal-to-noise ratio (SNR), especially at 1.5 T, where the impact of susceptibility artefacts is lower (52).

However, at 3.0 T, the use of spoiled gradient echo (GRE) could be a better option, despite being a slower technique. Another important aspect of the acquisition is that the MR signal can be affected by spatial variations due to differences in the coil sensitivity profiles; to reduce their effect on the data, a proton density (PD) image can be acquired without saturation preparation to normalise the signal (52,53). The normalisation process of dividing the T1-weighted images with contrast by the PD image can be performed by first smoothing the PD image to avoid pixel mismatch and noise amplification.

To reduce image degradation due to motion, pCMR acquisitions are typically performed with electrocardiogram (ECG) triggering and during a prolonged breath-hold (54). However, it may be difficult for patients to sustain respiration during the full first passage of the contrast bolus (around 40 seconds), particularly at stress. In this context, free-breathing acquisitions have become more frequent, especially because correcting regular and shallow motion may be easier than the rapid movement originated when the patient fails to maintain the breath-hold and gasps for air. Further information on motion correction strategies for pCMR can be found in **Section 6**.

Similarly to other quantification methodologies, partial volume effects can happen at the border of the myocardium. This may be relevant for both blood on the endocardium border and fat on the epicardium border. The contribution of fat to the perfusion measurements can be removed through fat suppression (19) or the use of multi-echo sequences. Multi-echo Dixon sequences additionally allow the acquisition of fat images that do not show contrast variations over time and can be used to avoid partial volume effects while correcting for motion, thus potentially improving image quality in free-breathing conditions (55,56).

Another limitation that should be considered in pCMR is the presence of dark rim artefacts (57), which appear when contrast reaches the left ventricular cavity before it appears in the myocardium, and that can be mistaken for true subendocardial perfusion defects. Dark rim artefacts may be caused by Gibbs ringing (58), susceptibility artefacts, motion (59), reduced spatial resolution, or partial volume effects (60). These artefacts can be distinguished from true defects when considering that they appear in the first perfusion images (first 5 R-R intervals) and have a range of one pixel, and by comparing stress and rest perfusion images. High-resolution pCMR can also be used to reduce dark rim artefacts and improve diagnostic accuracy (61,56).

# 5. Tracer-Kinetic Modelling: obtaining quantitative perfusion maps

Once the dynamic contrast-enhanced T1-weighted images have been acquired, the next goal in quantitative pCMR is to obtain quantitative measurements of myocardial blood flow. This requires converting the time-dependent MR signal intensity to contrast agent concentration before tracer kinetic models can be applied to extract quantitative measurements.

## 5.1. Conversion to Contrast Agent Concentration

It is often assumed that the observed contrast enhancement in the signal (the T1 relaxivity, $R_1 = 1/T_1$) during tracer passage is proportional to the change of contrast concentration in the myocardial tissue. This approximation holds true, although not perfectly linear, for short saturation delays or low contrast agent concentrations, as in the dual-saturation and dual-bolus sequences, respectively.

In more detail, the first step towards quantification is to convert the acquired signal intensity (T1-weighted) in the images into T1 variation. One possible assumption is that the measured MR signal is linearly proportional to the $R_1$ of that region; however, that approximation is only adequate for longer T1 values and can deviate for longer TS times and higher contrast agent doses (53). By knowing the MR signal model, for example, for GRE sequences, the pixel-wise normalised signal $S/S_{PD}$ can be described by an analytical expression of the signal evolution that can be applied to obtain the respective T1 value at each time-point. In addition, this allows correcting for surface coil variations.

Having estimated the T1 variation during tracer passage, the next step is to convert it into contrast agent concentration, according to the linear expression:

$$\frac{1}{T_1} = \frac{1}{T_1^{pre-contrast}} + r_{Gd}\,[Gd],$$

where $1/T_1^{pre-contrast}$ represents the native T1 relaxivity (pre-contrast), $r_{Gd}$ is the T1 relaxivity of the gadolinium-based contrast agent, and $[Gd]$ stands for the contrast agent concentration.

The linear equation above is a good description of the contrast agent concentration. However, the estimation of the T1 values for every dynamic is affected by the saturation pulses and the signal readout, resulting in a non-linear signal model, which becomes more pronounced as the contrast concentration increases, particularly affecting the AIF. Thus, the AIF kinetics differ from those of the myocardial tissue, requiring alternative acquisition schemes, as described in **Section 4**.

## 5.2. Tracer-Kinetic Modelling

The final step to obtain quantitative myocardial perfusion maps requires converting the contrast agent concentration into blood flow metrics. There are three main classes of methods that can be employed for this purpose: *model-independent deconvolution*, *compartmental models*, and *spatially distributed models*.

### 5.2.1. Model-independent deconvolution

*Model-independent deconvolution* methods do not require any assumption of the blood flow, but consider instead that the measured concentration variation $C_{tissue}$ (i.e. myocardial tissue response function) results from the fact that the arterial input $C_{AIF}$ is shaped by an impulse response function $h$:

$$C_{tissue}(t) = h(t) \otimes C_{AIF}(t),$$

where $\otimes$ denotes the convolution operation (62). Since the AIF and the tissue response are known, the blood flow in a specific region can be obtained through a deconvolution operation. The solutions are typically constrained according to prior physiological information so that the obtained impulse response is meaningful and suitable to be used as a perfusion indicator. A popular constraint in quantitative pCMR is to model the input response as a Fermi function with general parameters (without a direct physiological meaning) $A$, $\mu$ and $k$:

$$C_{tissue}(t) = \frac{A}{exp(\frac{t-\mu}{k}) + 1} \otimes C_{AIF}(t).$$

Through deconvolution, the parameters $A$, $\mu$ and $k$ are estimated and the myocardial blood flow is obtained as the maximum of the Fermi function.

### 5.2.2. Compartmental models

*Compartmental models* describe the exchange of contrast agent between blood and myocardial tissue by dividing the modelled region into a collection of interacting homogeneous compartments, such as the intravascular space and the interstitial (or extravascular extracellular) space, and defining how the tracer moves through and transverses permeable barriers between each compartment. These models assume that the concentration of contrast agent within a compartment at any given time is uniform and that the rate of change in contrast agent concentration is directly proportional to its concentration. Generally, models with a larger number of compartments can better capture the tissue physiology, i.e. are more realistic, and allow the estimation of several tracer-kinetic parameters, but at the expense of higher analysis complexity.

**Figure 6** shows schematic illustrations of six compartmental models. In pCMR, the modelled region (myocardial tissue) is typically divided into two compartments: the blood plasma, where the contrast agent is traversing, and the extravascular extracellular space, which receives contrast agent from the blood. To evaluate myocardial perfusion, the concentration of contrast agent in the blood, estimated with the AIF, is summed to the estimated contrast agent concentration in the extravascular extracellular space.

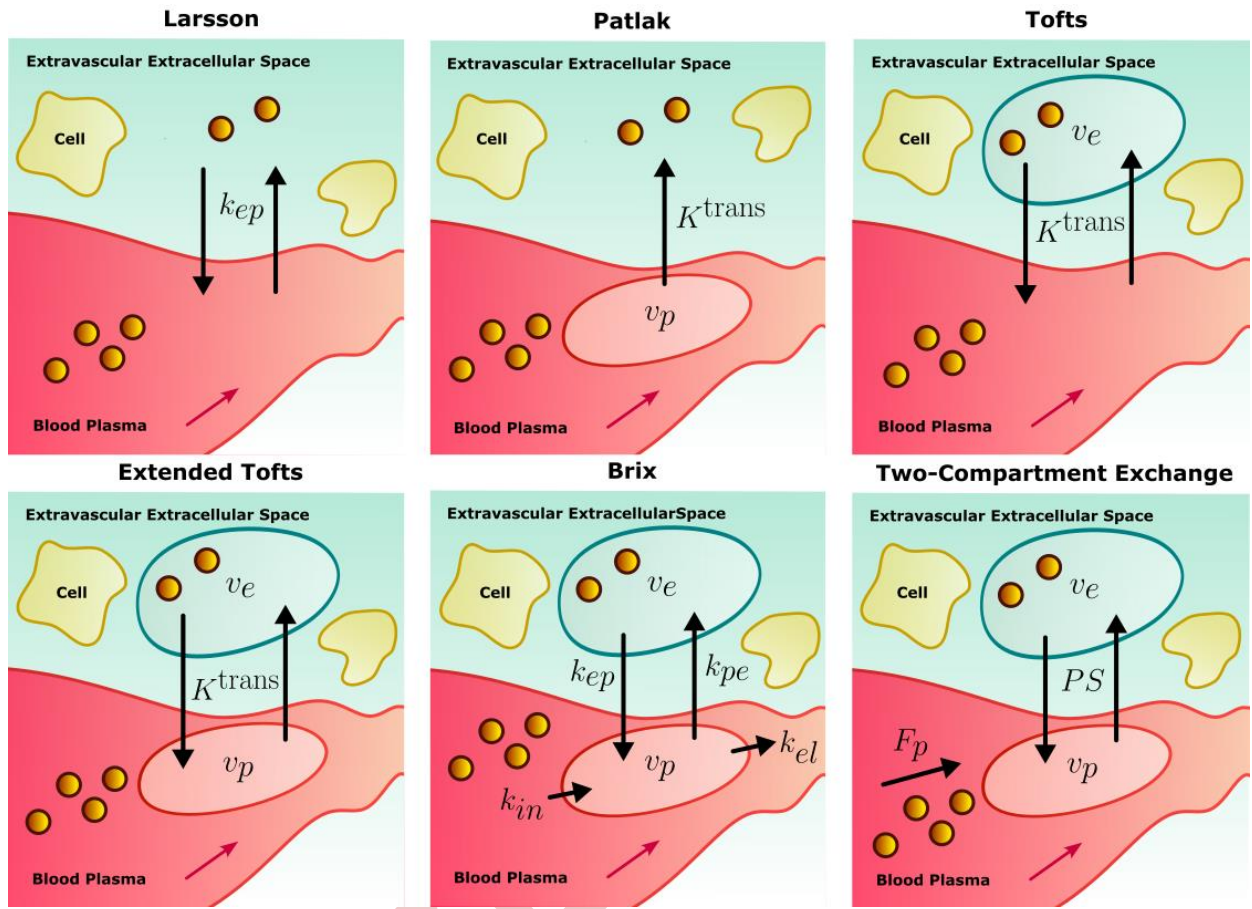

**Figure 6. Illustrations of compartmental models for perfusion analysis.** Each compartmental model defines a set of parameters to estimate that describe the transfer of the contrast agent from the blood to the extracellular extravascular space ($K^{trans}$, $k_{pe}$), from the extracellular extravascular space to the blood ($k_{ep}$), and in both directions ($k_{ep}$, $K^{trans}$, $PS$); the volumes of the blood compartment ($v_p$) and of the extracellular extravascular space ($v_e$); the blood flow ($F_p$); and the rates of administration ($k_{in}$) and elimination ($k_{el}$) of the contrast agent in the blood.

One of the earliest compartmental models to be developed was the Larsson model (63), which considers two compartments: the blood plasma in the capillary and the extravascular extracellular space. The Larsson model assumes that the tracer moves through both compartments according to a single rate constant $k_{ep}$, combining information from the capillary blood flow, the extraction fraction and the fractional extravascular extracellular space. The tracer concentration in the blood plasma, $C_{blood}(t)$, is calculated according to the AIF, which is estimated by fitting a sum of three exponentials with auxiliary variables $a_i$ and $m_i$, which results in the following tracer concentration on the extravascular extracellular compartment:

$$C_{tissue}(t) = k_{ep} \sum_{i=1}^{3} \frac{a_i}{m_i - k_{ep}} (exp(-k_{ep}t) - exp(-m_i t)).$$

The Larsson model is only applicable to situations where the tissue permeability is considerably lower than the flow, but the single rate constant can be easily estimated via numerical optimisation.

The Patlak model (64) considers solely the transfer of the contrast agent from the intravascular space to the extracellular extravascular space, whilst disregarding the inverse direction. This is expressed in the following form:

$$C_{tissue}(t) = K^{trans} \otimes C_{blood}(t) + v_p C_{blood}(t),$$

where $K^{trans}$ is the volume transfer coefficient, which is related to the plasma blood flow and the permeability surface-area product $PS$, and $v_p$ is the intravascular volume fraction. With this representation, there is a linear relationship between the signal curve in the tissue and in the plasma (i.e. AIF), which simplifies calculations. However, since the passage back to the vasculature is only negligible during the first initial part of the tracer's arrival to the myocardium, the model is only a good approximation in that instance. Furthermore, the model cannot estimate the plasma flow and the tissue permeability separately, since both parameters are included in $K^{trans}$.

The Tofts model (65) considers that the tracer concentration in the tissue is given only by the extravascular extracellular compartment, with volume fraction $v_e$, and disregards the blood plasma contribution. According to this model, the contrast agent moves between the blood and the extravascular extracellular space at a rate determined by the transfer constant $K^{trans} = k_{pe}/v_e$. The concentration of contrast agent in the blood $C_{blood}(t)$ is given by the AIF, and the tracer concentration in the tissue is then given by:

$$C_{tissue}(t) = K^{trans} exp\left(-\frac{K^{trans}}{v_e} t\right) \otimes C_{blood}(t).$$

The Tofts model is an adequate choice when the tissue is weakly vascularized. The extended Tofts model is an extension of this method that also includes the intravascular volume fraction through an additional parameter $v_p$, allowing its application to more perfused systems (66):

$$C_{tissue}(t) = K^{trans} exp\left(-\frac{K^{trans}}{v_e}t\right) \otimes C_{blood}(t) + v_p C_{blood}(t).$$

However, like the Patlak model, neither of these models can estimate the plasma flow and the tissue permeability separately, since both are included in $K^{trans}$.

Similarly to the Extended Tofts model, the Brix model (67) considers both the intravascular and the extravascular extracellular compartments, with respective volume fractions $v_p$ and $v_e$. However, the latter admits a forward rate constant $k_{pe}$ and a reverse rate constant $k_{ep}$. In addition, the administration and elimination of the contrast from the plasma are also considered through the constants $k_{in}$ and $k_{el}$, respectively. The contrast agent concentrations are modelled as follows:

$$C_{tissue}(t) = \frac{k_{in}k_{pe}}{v_p(k_{ep}-k_{el})}\left(\frac{exp(-k_{el}t)}{k_{el}}(exp(k_{el}t')-1) - \frac{exp(-k_{el}t)}{k_{ep}}(exp(k_{ep}t')-1)\right),$$

$$C_{blood}(t) = \frac{k_{in}}{v_p k_{el}}(exp(k_{el}t')-1)exp(k_{el}t),$$

$$t' = t, if\ 0 \leq t \leq \tau$$
$$t' = \tau, if\ \tau \leq t$$

where $\tau$ is the time-span over which the contrast is administered. When fitting to the measured signals, the Brix model only requires estimation of three parameters: $k_{ep}$, $k_{el}$ and an arbitrary constant $A^{Brix}$ that depends on the tissue properties and on the MR sequence.

The Two-Compartment Exchange model (68) includes the extravascular extracellular space, the intravascular space, and describes the tracer exchange between them over time. The model considers that the contrast agent exchanges between compartments occur at a symmetric rate of $PS$, and that the tracer concentrations in each compartment are described by the following differential equations:

$$\frac{dC_{blood}(t)}{dt} = \frac{PS}{v_p}(C_e(t) - C_{blood}(t)) + \frac{F_p}{v_p}(C_{AIF}(t) - C_{blood}(t)),$$

$$\frac{dC_e(t)}{dt} = \frac{PS}{v_e}(C_{blood}(t) - C_e(t)),$$

$$C_{tissue}(t) = v_p C_{blood}(t) + v_e C_e(t),$$

where $C_e(t)$ is the extravascular extracellular space concentration, $PS$ is the permeability-surface area product, and $F_p$ is the flow experienced by the blood plasma. On the one hand, calculations for the Two-Compartment Exchange model are more complex since it requires solving a pair of coupled ordinary differential equations; on the other, the model allows separate estimates of permeability and plasma blood flow.

### 5.2.3. Spatially distributed models

Spatially distributed models do not assume homogeneous contrast agent concentration in compartments, but instead assume that there are gradients in contrast agent concentration along the flow direction. Thus, unlike compartmental models, spatially distributed models account for both spatial and temporal variations of the contrast agent, so they are expected to reflect the underlying physiology more accurately. However, the complexity of these methods and high computational requirements often make them impractical to deploy in many clinical scenarios. Popular spatially distributed models include the Tissue Homogeneity model (69), the Distributed Parameter model (70) and the Blood Tissue Exchange model (71) (**Figure 7**).

The Tissue Homogeneity model (69) is an extension of the Two-Compartment Exchange model which models the contrast agent concentration as a function of distance along the capillary as well. This model assumes that, in addition to the conditions specified by the Two-Compartment Exchange model, one needs to consider that, in the intravascular compartment, the tracer concentration varies along the direction parallel to the capillary and that the particles along the capillary all travel with the same velocity (*plug-flow* model). The extravascular extracellular space is assumed to be a single compartment without contrast agent concentration gradients, unlike the intravascular space, described as a flow system. The model is characterised by the set of differential equations below:

$$v_p \frac{\partial C_{blood}(x,t)}{\partial t} = -PS(C_{blood}(x,t) - C_e(t)) - L\, F_p \frac{\partial C_{blood}(x,t)}{\partial x},$$

$$v_e \frac{\partial C_e(t)}{\partial t} = PS(C_{blood}(t) - C_e(t)),$$

$$C_{tissue}(t) = v_p C_{blood}(t) + v_e C_e(t),$$

where $L$ is the capillary length.

The Distributed Parameter model (70) is a generalisation of the Tissue Homogeneity model which considers that the tracer concentration is also changing in the extravascular extracellular compartment, but that there is no exchange or transportation of contrast agent in this compartment along the capillary direction. The model is described by the differential equations:

$$v_p \frac{\partial C_{blood}(x,t)}{\partial t} = -PS(C_{blood}(x,t) - C_e(x,t)) - L\, F_p \frac{\partial C_{blood}(x,t)}{\partial x},$$

$$v_e \frac{\partial C_e(x,t)}{\partial t} = PS(C_{blood}(x,t) - C_e(x,t)),$$

$$C_{tissue}(t) = v_p C_{blood}(t) + v_e C_e(t).$$

A popular spatially distributed model for pCMR is the Blood Tissue Exchange model (BTEX) (71). The BTEX model introduces the concept of a capillary unit as a set of endothelial cells, interstitial fluid, parenchymal cells, and plasma. Each element of the capillary unit is defined by an equation that describes the concentration of contrast agent as a function of time and space, typically containing as variables the permeability-surface area product, the plasma flow, and the capillary and cell volumes. Then, a multi-capillary model with capillaries having different flows can be considered as well.

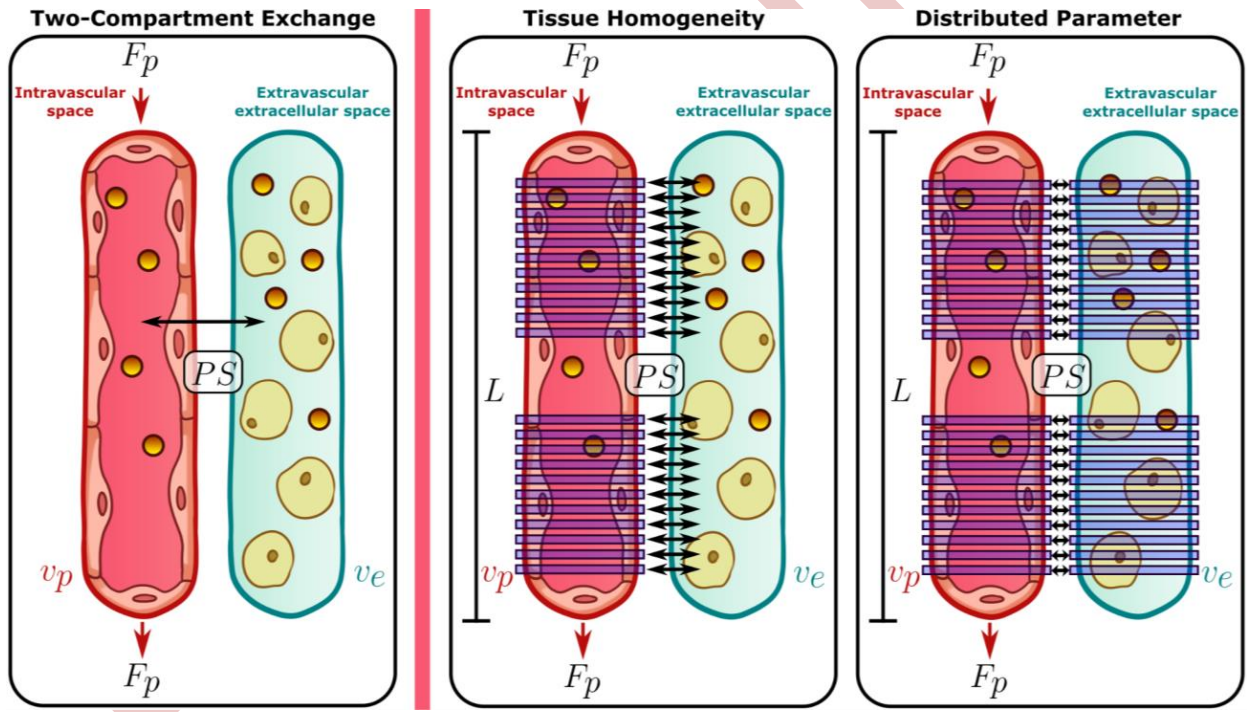

**Figure 7. Diagram depicting the exchange of contrast between the intravascular space and the extravascular extracellular space, as described by popular spatially distributed models for perfusion analysis and compared to the Two-Compartment Exchange model.** Fp describes the direction of the blood flow; vp and ve describe the volumes of the blood compartment and of the extravascular extracellular compartment, respectively; L is the length of the capillary; PS is the permeability-surface area product. Spatially distributed models assume the contrast agent concentration varies along the spatial dimension as well, and model this by considering

infinitesimal volumes over the intravascular and extravascular extracellular compartments; both models consider that the tracer concentration varies along the blood compartment; the Distributed Parameter model additionally considers that the tracer varies along the extravascular extracellular space.

# 6. Accelerated pCMR: imaging with high resolution and/or heart coverage

In pCMR, there is an inherent trade-off between cardiac coverage, spatial and temporal resolution, and SNR. For example, imaging at a higher spatial resolution using standard pCMR protocols results in reduced SNR or scan times. For this reason, when planning a clinical pCMR protocol, taking this trade-off into account is of paramount importance. Moreover, acquisitions typically require synchronisation with the cardiac and respiratory cycles to avoid motion-related artefacts, which comes at the expense of spatial resolution and heart coverage (72). A typical pCMR protocol can achieve an in-plane spatial resolution of approximately 2.5 mm for 3-4 short-axis slices. Acceleration techniques thus become crucial to overcome the limitations imposed by the trade-off between in-plane spatial resolution and coverage of the left ventricle.

## 6.1. Reconstruction of accelerated acquisitions

Different acceleration strategies are usually jointly applied in pCMR to improve resolution and coverage. Among these, one of the main acceleration sources consists in reducing the number of k-t samples required for the reconstruction (i.e., sampling below Nyquist limit or undersampling the k-space over time). Common k-space undersampling trajectories include Cartesian, Poisson disk, Cartesian pseudo-spirals, non-Cartesian radial and spiral, stack-of-stars (3D radial), stack-of-spirals (3D spiral), koosh-ball and spiral phyllotaxis (3D non-Cartesian) trajectories, amongst others (72). Then, the undersampled acquisition should be followed by a constrained reconstruction method to avoid aliasing artefacts in the reconstructed image. In MRI, such techniques include sensitivity encoding (SENSE) (73), compressed sensing (CS) (74), low-rank (LR) matrix completion (75,76), low-rank plus sparse (L+S) reconstruction (77), model-based reconstruction (78), and k-t principal component analysis (PCA) (79).

SENSE is a parallel imaging technique, where multiple-coil arrays are used for spatial encoding, compensating for the missing k-space information due to undersampling. Calibration of coil sensitivities is required, either acquiring a low-resolution image prior to the main scan or using auto-calibration strategies.

CS is based upon the principle that an image with a sparse representation in a known transformed domain can be recovered from randomly undersampled k-space data using nonlinear reconstruction techniques. The three fundamental requirements for the successful application of CS are: (a) the image must have a sparse representation in some transformed domain; (b) the undersampling pattern used must lead to incoherent or noise-like artefacts; and (c) a non-linear reconstruction algorithm, such as the Fast Iterative Shrinkage/Thresholding Algorithm (FISTA) or the Alternating Direction Method of Multipliers (ADMM), must be used to recover the image from the undersampled measurements. pCMR is a natural candidate for CS acceleration since the signal contains significant redundancy throughout the temporal domain and the required incoherence can be effectively accomplished by k-t (pseudo-) random undersampling.

LR is a technique that formulates the image reconstruction as a low-rank matrix recovery problem. In pCMR, there are strong temporal-correlations, so the number of degrees of freedom in the reconstruction can be constrained with LR methods (75). LR thus extends CS reconstruction to matrices and finds missing or corrupted entries under low-rank and incoherence conditions (76). L+S reconstruction consists in the combination of CS and LR and attempts to find a solution that is both low-rank and sparse. More specifically, L+S decomposes pCMR images into a low-rank component L that captures the temporally-correlated background and a sparse component S that captures the dynamic contrast-enhancement.

Model-based methods incorporate physics models into the image reconstruction problem to enable the prediction of quantitative parameters directly from undersampled data (80–82). Therefore, in quantitative pCMR, this method avoids all the intermediate steps, including the reconstruction of the dynamic pCMR images and conversion from MR signal to contrast agent concentration.

Finally, k-t PCA relies on the assumption that the acquired k-space data will have a high degree of spatiotemporal correlations, hence it can be represented in a compact representation. PCA is a tool that extracts the main features of the acquired data (principal components) to form an orthonormal basis suitable to represent the data in a lower dimensional space. It can be used as a dimensionality reduction transform in CS or LR, saving processing time and power requirements. With k-t PCA, k-space is sparsely sampled and the resulting representation in the lower dimensional space will have little overlap of the signal replicas, allowing the recovery of the original signal.

Specifically, as shown in Table 1 (column *Recon method*), k-t SENSE (83), k-t PCA (84,85), CS (86–91), LR (92,93), L+S (56,77,94) and model-based reconstruction methods (82) have been proposed over the years to accelerate pCMR acquisitions. In general, it can be noticed that the methods that achieve high in-plane spatial resolution ($< 2\times2$ mm$^2$) are limited in terms of anatomical coverage and vice-versa.

Three-dimensional (3D) acquisitions have also been developed to achieve full heart coverage (84–86,91,93,95). However, these methods are often constrained by limited in-plane spatial resolution ($> 2\times2$ mm$^2$) due to the need for a sufficiently short readout to reduce the impact of cardiac motion and dynamic contrast changes. Alternatively, simultaneous multi-slice (SMS) sequences have recently been proposed combined with undersampling techniques to increase coverage without sacrificing in-plane spatial resolution (87,88,90,96). SMS is a 2D imaging method that utilises multiband RF pulses to simultaneously excite multiple slices. Thus, several slices can be imaged in the same amount of time as a traditional single-slice acquisition.

## 6.2. Handling respiratory and cardiac motion

In conventional pCMR, data acquisition is typically synchronised with the ECG signal, thereby ensuring that images in each slice are acquired during the same cardiac phase. Consequently, image quality relies on the regularity and precision of the ECG triggers. However, ECG is often contaminated in the MRI environment due to magnetohydrodynamic effects. Furthermore, in subjects with arrhythmias, variations in the R-R interval width are challenging for accurate ECG gating. This is of particular importance in stress scans, in which heart rate changes are common even in relatively healthy patients. These issues can result in quality degradation and non-diagnostic images. To address the problem of poor gating, several free-running methods (non-ECG-gated) have been proposed, with continuous data acquisition workflows; see Table 1 (column *Free-running*). In this case, due to the presence of cardiac motion, the ungated dataset must be retrospectively sorted to create self-gated perfusion images.

Regarding respiratory motion, pCMR acquisitions have traditionally been performed during breath-hold conditions, but a growing number of reconstruction schemes include strategies that are robust to free-breathing acquisitions (see Table 1, column *Free-breathing*). Under free-breathing conditions, the application of motion estimation and motion compensation algorithms becomes necessary to reconstruct high quality images without motion artefacts (56,82,85,93,96). In addition, it is not uncommon to apply motion estimation and motion compensation to breath-hold acquisitions to compensate for residual respiratory motion caused by imperfect breath-holding (89,96).

Different approaches for motion estimation/compensation applied to pCMR have been reported. Typically, the respiratory motion estimation step is performed in the image domain by means of different image registration algorithms, following either a pairwise (82,85,89,96) or a groupwise (93) approach. This process is carried out only once to obtain the corresponding deformation fields, modelled as rigid (56,82,97) or non-rigid (85,90,93) transformations, prior to the motion-compensated reconstruction. In contrast, Chen *et al.* (89) proposed BLOSM, a scheme where motion corrected image regions are included in the LR formulation to increase sparsity. Thus, the local motion fields are iteratively updated after 50 iterations of the reconstruction following a coarse-to-fine strategy; first with a rigid and later with a non-rigid registration.

Different motion compensation strategies have also been proposed. In some cases, the respiratory motion is modelled as an in-plane rigid translation, and it is corrected as a linear phase shift in the k-space domain (56,82). In other cases, a motion compensation operator that aligns the images is defined and included in the reconstruction problem, either in the data consistency term (93,96) or in the regularisation term (85,85,90).

**Table 1.** Overview of technical details of the pCMR methods in order of publication date.

| Ref. | Recon. method | Acquisition | AF | #Slices | Resolution [mm³] | Acq. Window [ms] | Stress | Scanner field strength [T] | Free-breathing | Free-running |
|---|---|---|---|---|---|---|---|---|---|---|
| Vitanis, 2011 (84) | k-t PCA | 3D Cartesian | 7 | 10 | 2.3×2.3×10 | 225 | Yes | 3 | No | No |
| Chen, 2012 (86) | CS | 3D Radial (Stack-of-Stars) | 9 | 8 - 10 | (1.8-2.8)×(1.8-2.8)×(6-10) | 300 | No | 3 | No | No |
| Shin, 2013 (95) | k-t SENSE | 3D Radial (Stack-of-Spirals) | 5 | 10 | 2,4x2,4x9 | 230 | No | 1.5 | No | No |
| Harrison, 2013 *(98) | CS | 2D Radial | Un | 4 - 5 | 2×2×8 | - | Yes | 3 | Yes | Yes |
| Chen, 2014 * (89) | LR | 2D Cartesian | 4** | 1 – 4 | (2.9-2)×(2.9-2.7)×8 | Un | No | 1.5 | No | No |
| Stäb, 2014 (87) | CS | 2D SMS Cartesian | 5 (SMS = 2) | 6 | 2.0×2.0×8 | 223.4 | No | 3 | No | No |
| Schmidt, 2014 * (85) | k-t PCA | 3D Cartesian | 10 | 10 | 2.3×2.3×10.0 | 205 - 225 | No | 3 | Yes | No |
| Akçakaya, 2014 (91) | CS | 3D Cartesian | 10 | 6 | 2.3×2.3×10.0 | 250 | No | 1.5 | Yes | No |
| Otazo, 2015 (77) | L + S | 2D Cartesian | 8 | 10 | 1.67×1.67×8 | Un | No | 3 | No | No |
| Sharif, 2015 (99) | CS | 2D Radial | 12 | 3 | 1.4×1.4×10 | - | Yes | 3 | No | Yes |
| Likhite, 2016 * (100) | CS | 2D Radial | Un | 4 | 1.83×1.83×8 | - | Yes | 3 | Yes | Yes |
| Christodoulou, 2018 (92) | LR | 2D Radial | | 1 | 1.7×1.7×8 | Un | No | 3 | No | Yes |
| Correia, 2019 * (82) | Model-based | 2D Radial | 20, 30 and 40** | 1 | 2.8×2.8×10 | 224.3 | No | 3 | Yes | No |
| Tian, 2019 * (90) | CS | 2D SMS Radial | Un | 6-18 (2-6 SMS sets/3 | 1.8×1.8×8 | Un | Yes | 3 | Yes | Yes |

| | | | | slices each) | | | | | | |
|---|---|---|---|---|---|---|---|---|---|---|
| Tian, 2020 * (101) | LR | 2D SMS Radial | Un | 9 - 12 | 1.8×1.8×5-8 | - | Yes | 3 | Yes | Yes |
| McElroy, 2020 (88) | CS | 2D SMS Cartesian | 11 (SMS=2; R=5.5) | 6 (2 patients, full left ventricle) | 1.4×1.4×10 | 137 | No | 1.5 | No | No |
| Wang, 2021* (96) | L1-SPIRiT | 2D SMS Spiral | SMS=2-3 | 6 or 8/ 6 or 9 | 1.25×1.25 × 10 | Un | No | 3 | No | No |
| Sun, 2022 (94) | L + S | 2D SMS Cartesian | 6 or 9 (SMS=3, R=2-3) | 9 | 1.5×1.5×(8-10) | 169 - 195 | No | 1.5 | No | No |
| Hoh, 2022 * (93) | LR | 3D Cartesian Pseudo-spiral | 10 | Full left ventricle | 2.5×2.5×10 | 240 | Yes | 1.5 | Yes | No |
| Tourais, 2022 * (56) | L+S | 2D Cartesian | 8 | 3 | 1.6×1.6×10 | 65.4 | No | 3 | Yes | No |

*Includes motion compensation. **Retrospectively undersampled. Un stands for unknown; other abbreviations as defined in the text.

A more detailed description of several recent methods is provided below. Sun *et al*. (94) extended the L+S idea originally described by Otazo *et al*. (77), to the SMS case to perform a joint multislice (and multiframe) reconstruction that enforces consistency with the acquired SMS data while enforcing temporal low-rank and sparsity within each of the multiple slices. In this work, however, motion was not accounted for.

Correia et al. proposed DIREQT (82), a model-based image reconstruction framework that combines image reconstruction with tracer kinetic modelling, to directly estimate quantitative pCMR maps from undersampled k-space data. Here, respiratory translational motion correction was performed in k-space before reconstruction. More recently, Hoh et al. (93) proposed a motion-informed locally LR reconstruction (MI-LLR). In this case, the authors tackled a 3D + t free-breathing problem and the solution was posed as a two-stage optimisation process. The first stage is an image reconstruction procedure followed by an image registration process across T dynamics that minimises the nuclear dissimilarity metric of warped images. The resulting motion fields map images into a common reference frame. Hence, in the second stage these fields are inverted so that the images that will be reconstructed in this common reference frame are mapped into their original k-spaces. The method is referred to as LLR since the regularisation terms include the nuclear norm of patches whose locations are randomly selected in each step of the optimisation to avoid blocking effects in the final solution. Respiratory motion was also corrected in two steps of the quantitative pCMR pipeline described in Tourais *et al* (56). This method uses a dual-echo Dixon acquisition to generate water- and fat-only images from undersampled k-t data. First, fat-only images, which are unaffected by the dynamic contrast-enhancement, are used to efficiently estimate rigid motion. Then, this information is used to rigidly correct the dual-echo k-space data. Finally, high-resolution motion-corrected water-only images are generated with a low-rank and sparsity constrained reconstruction. Non-rigid motion correction was then applied to water-only images as a fine-tuning process prior to myocardial perfusion quantification.

Several other MRI motion correction approaches have been recently proposed, which could potentially be applied to pCMR. For example, Huttinga et al. (102) proposed MR-MOTUS, a framework that enables the estimation of non-rigid motion from undersampled k-space data. MR-MOTUS assumes the availability of one image acquired at some point along the interval to be reconstructed; the measurements at other time instants are modelled by means of a Fourier transform in which the deformation fields are included in the complex exponential. The deformation fields are then obtained by inverting the model. This idea is further developed in (103) and (104) to incorporate a low-rank model for motion. An application to time-resolved cardiac imaging with/o contrast enhancement has also been reported (105). In this case, the MR-MOTUS framework was combined with the L+S model (106) to accommodate time-dependent contrast changes. The solution iterates between solving for the deformation field and then for the images.

# 7. Artificial Intelligence in pCMR: what value does it bring?

Improving pCMR to obtain higher temporal and spatial resolution and heart coverage in a seamless workflow requires a cost-effective and time-efficient strategy for integration into routine clinical practice. Artificial Intelligence (AI) can have a significant role in addressing the current pCMR challenges, namely to optimise image acquisition protocols, accelerate scans, improve image post-processing and reporting, and incorporate these in clinical decision-making models.

## 7.1. Basic concepts of artificial intelligence

AI is defined as a set of techniques that machines use to mimic the way humans solve problems or make decisions. A sub-branch of AI, *machine learning*, uses algorithms and statistical models to find patterns and make extrapolations from large amounts of data. Machine learning algorithms can be classified into several categories, two of which are supervised and unsupervised learning methods (107). In *supervised learning*, the most commonly employed form of machine learning, training datasets contain labels designed by experts (e.g., pathological status, segmentation masks, or quantitative measurements), which are used to guide the algorithm optimisation to a desirable solution. In *unsupervised learning*, training datasets lack labels, and the algorithms aim instead to learn patterns from the data through clustering and association. Unsupervised learning requires minimum human supervision and is a useful strategy for applications where a large database of labelled data is challenging, or even impossible, to curate. *Self-supervised learning* is a subset of unsupervised learning where labels are generated from the given data itself; it can be used to reconstruct images from undersampled k-space data by either enforcing data consistency using the physical model in the reconstruction (physics-guided network), or by training networks with a subset of the undersampled data acquired (108).

The most common applications of machine learning in MRI are part of a class of methods called *deep learning* (DL) strategies, where neural networks with multiple layers are employed. *Neural networks* consist of a set of placeholders for numerical expressions called *neurons*, which produce an output by applying that expression to a given input (**Figure 8-a**) and are grouped into network layers. Implementing a neural network requires defining the network topology and performing three steps: training, validation, and testing.

During the training stage, a typically-large amount of data (i.e., *training dataset*) is progressively processed and fine-tuned through all the neurons, thus extracting meaningful features that contribute to the final decision. The training stage is accomplished by dividing the training dataset into smaller subsets of data, called *batches*, and iteratively updating the network parameters, called *network weights*, once each batch has been processed. To update the network weights, a *loss function*, which measures the network performance on the current batch, must be applied. The loss function instructs the network on how its weights should be updated. Weights are updated through an algorithm called *backpropagation*, where the gradients of the loss function with respect to the network weights are calculated and used as a measure of intensity and direction that each weight should change to make better predictions from the data. Training achieves one *epoch* once the full training dataset has been seen by the network; typically, networks train for multiple epochs until reaching convergence.

Networks are usually trained for varying combinations of *hyperparameters*, which can include the network architecture and optimisation variables such as the loss function and the batch size. The optimal configuration of hyperparameters is chosen during the validation stage, where each network is applied to a generally smaller distinct amount of data, called *validation dataset*. The outputs of these networks on the validation dataset are compared to each other, allowing the user to pick the optimal combination of hyperparameters. Once this has been done, the chosen network can be applied to the *testing dataset*, which consists of new unseen data. The validation dataset is also crucial to avoid *overfitting*, i.e., that the network learns the training dataset exceptionally well, losing the ability to generalise on unseen data. Additionally, networks can be trained with *regularisation* to avoid overfitting. Regularisation comprises a range of methods to increase a model's generalizability, typically at the sake of increasing the training error, which include additional terms in the loss function, adding new artificial samples to the training dataset (*data augmentation*) and limiting the number of epochs during training.

## 7.2. Building blocks of a neural network

Convolutional neural networks (CNNs) are a class of DL networks designed to work with images that learn multiple features for a given input. A typical CNN is composed of multiple layers (**Figure 8-b**). *Convolution layers* employ a set of filters that are applied to the image to produce spatially dependent features (filter maps) for the next layer. *Pooling layers*, such as max and average pooling, downsample feature maps by reducing their spatial dimensions, which eases the computational demand of training and testing the network. *Skip layers* connect the feature maps of non consecutive layers, avoiding gradients becoming too small during backpropagation (i.e., gradient vanishing) and allowing the propagation of fine image features as the number of layers increases; the layers between the skip layers form a *residual block*. *Batch normalisation layers*, as well as normalised inputs and outputs, avoid gradient vanishing and thus stabilise the training process.

The most popular CNN architecture used in MRI is the UNet (**Figure 8-e**). The UNet contains an encoder path, consisting of convolutional and max-pooling layers that reduce the spatial information in search of relevant features, and a decoder path, consisting of a series of up-sampling convolutional layers that recover the original spatial dimensions. The encoder and decoder are connected through long-range skip connections.

Recurrent Neural Networks (RNNs) (**Figure 8-c**) are another class of neural networks developed for analysing sequences of data and which are commonly used for modelling the temporal dependencies present in time series, sequences or videos. Variants of RNNs include the Long-Short Term Memory RNN (LSTM) and the Gated Recurrent Unit (GRU), which were developed to improve handling of long-term dependencies. RNNs are a common block of unrolled architectures (**Figure 8-d**) for image reconstruction, where parts of the traditional iterative optimisation strategies are replaced with neural networks and each step of the optimisation process is learnt.

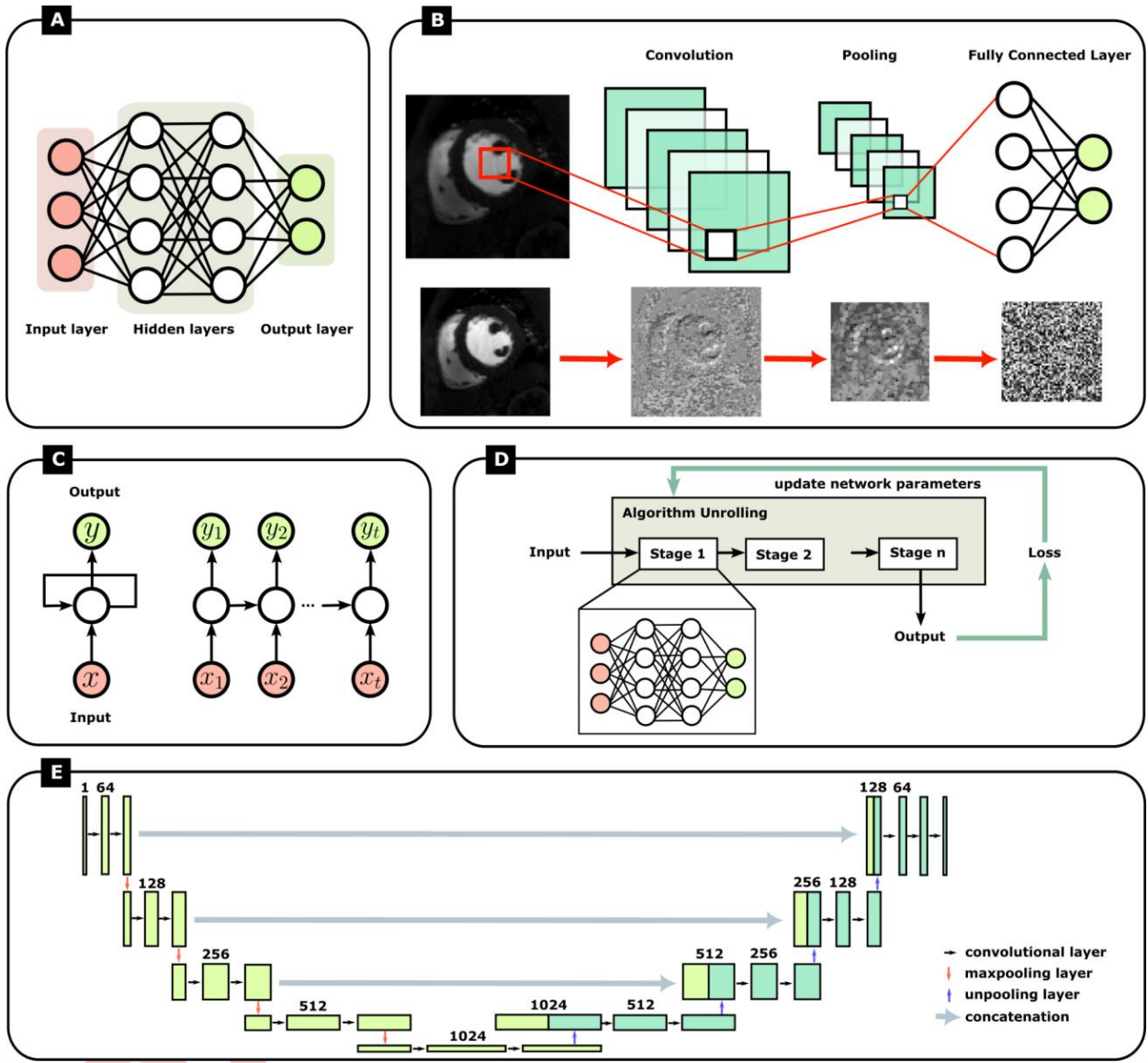

**Figure 8. Deep learning architectures.** (**a**) General framework of a neural network: the inputs are processed by an input layer, followed by a number of hidden layers that learn patterns in the training data, and an output layer that produces a decision. (**b**) Typical structure of a CNN. Example of the feature maps obtained on an image by applying a convolutional layer, a maxpooling layer and a fully-connected layer that applies a linear operation. (**c**) Typical structure of an RNN. (**d**) Algorithm unrolling in physics-driven deep learning methods. An iterative algorithm is unrolled for n iterations or stages and trained end-to-end. The network parameters can be shared or vary across stages. (**e**) Example of the UNet architecture.

## 7.3. Applications

Multiple deep learning architectures have been developed to facilitate the quantitative pCMR workflow. The most common application is to train a network to resolve artefacts from undersampled k-space in order to allow faster acquisitions with the same image quality. Multiple works have employed this strategy to reconstruct CMR images (108–114); most deep learning algorithms for pCMR image reconstruction specifically consist of supervised UNet adaptations (115,116,97), but self-supervised approaches have also been proposed to combat the lack of labelled data (117,118). Architectures have also been developed to facilitate motion-correction in CMR, and combine motion-correction with reconstruction. Additionally, multiple UNet-based DL applications have been proposed to perform automatic segmentation of the heart regions (119–124). Finally, a model has been proposed to convert signal intensity into contrast agent concentration and calculate MBF maps (125), and another to predict CAD from image and other clinical information (126).

### 7.3.1. Image Reconstruction

Accelerating pCMR image acquisition through k-space undersampling introduces aliasing artefacts in the image domain that can be resolved by solving an iterative regularised optimisation problem. Traditional means for image reconstruction from undersampled data, such as CS and LR matrix completion, are typically slow and require careful selection of optimisation parameters and regularisation functions. DL-based approaches are an alternative way of resolving aliasing artefacts and solving general inverse problems, which have enabled higher acceleration factors while offering extremely fast reconstruction speeds.

Most DL methods developed for pCMR consist of supervised UNet adaptations. Fan et al (115) proposed a UNet architecture to reconstruct 6.4-fold accelerated radial-undersampled pCMR k-space data with the same image quality as CS but with a 10-fold faster computation time. The network was trained using coil-combined zero-filled images as inputs and the corresponding CS reconstructions as outputs. Le et al. (116) proposed the 3D Residual Booster UNet, an adaption of the UNet for processing data with temporal dependencies, to reconstruct radial SMS myocardial perfusion data. This network was implemented by combining two 3D UNets into a booster network framework to compensate for the weaknesses of each network, producing a stronger learner, while including an additional residual framework to improve gradient flow. With this method, reconstructing one radial SMS dataset required approximately 8 seconds, 200 times faster than CS. The DEep learning-based rapid Spiral Image REconstruction (DESIRE) (97) consists of a 3D UNet that takes as input complex-valued data, where the real and imaginary parts are concatenated along the channel dimension, developed for reconstructing high-resolution spiral pCMR myocardial scans for SMS acquisitions, achieving an in-plane acceleration of 5-fold.

Self-supervised networks for reconstruction have been proposed to address the lack of labelled ground truth information, i.e. fully-sampled pCMR data. Martìn-González et al. (117) developed the SElf-supervised aCcelerated REconsTruction (SECRET) method, a self-supervised physics-informed CNN, to accelerate retrospectively radial-undersampled pCMR scans, which acts as a k-space "denoiser". First, the k-space data is transformed into the image domain, resulting in an aliased image that enters a CNN tasked with removing undersampling artefacts. The reconstructed images are then transformed back to k-space following the acquisition model and compared to the measured data to optimise the network. SECRET makes use of skip connections and a residual block to improve image reconstruction, and provides high-quality reconstructed images from up to 10× undersampled data in 0.15s. Demirel et al. (118) presented a physics-guided deep learning strategy for reconstructing free-breathing 3-fold accelerated SMS myocardial pCMR based on the self-supervised learning via data undersampling (SSDU) method, where networks are trained with a subsample of the acquired undersampled k-space, whilst the other subsample is used to compute the loss function. A signal intensity informed multi-coil operator was added to the SSDU to capture the signal intensity variations across time-frames so that the network could output images with a uniform contrast, thus achieving higher generalisability across time-frames and patients.

While AI models specially applied to pCMR reconstruction from undersampled data are still limited, there is a vast amount of literature on DL reconstruction tools for CINE imaging which could be adapted to pCMR. Aggarwal et al. (109) introduced a Model Based Deep Learning Architecture for Inverse Problems (MoDL), an end-to-end physics-guided image reconstruction framework with a CNN based regularisation prior. MoDL consists of interleaved CNN blocks that capture the information on the image set, and blocks that enforce data consistency by applying the conjugate gradient algorithm, with shared weights across iterations. Biswas et al. (110) developed MoDL-STORM, a MoDL variant containing image regularisation penalties to reconstruct free-breathing and free-running CMR data from highly undersampled multi-channel measurements. The loss function was formulated as the sum of the data consistency term, the CNN denoising prior, and a SmooThness Regularization on Manifolds (SToRM) prior that exploits the nonlocal redundancies between images in the dataset, which are specific to the cardiac and respiratory patterns of the subject. While MoDL cannot reconstruct 50-fold accelerated images, a high undersampling factor necessary to enable free-breathing ungated acquisitions, MoDL-STORM was able to retain image quality in this setting.

Qin et al. (111) proposed a Convolutional Recurrent Neural Network (CRNN) to reconstruct high quality CMR images by jointly exploiting temporal dependencies of dynamic MR sequences as well as the iterative nature of traditional optimisation algorithms. The CRNN-MRI method consisted of a CRNN block followed by a data consistency layer. The CRNN block contained five units: a bidirectional convolutional recurrent unit that exploited the temporal dependency of the dynamic sequences, three CRNN layers and one CNN unit. CRNN-MRI enabled reconstruction from 9x retrospectively Cartesian undersampled CINE data in 3s.

Küster et al. (112) developed CINENet to reconstruct multi-coil complex-valued 4D (3D + time), trained with retrospectively 3D Cartesian cardiac CINE data with undersampling factors up to 8-fold. CINENet consisted of a cascade of 4D UNets modified to perform unrolled optimisation with complex-valued convolutions and activation functions, alternated with data consistency blocks. CINENet allowed for 3D isotropic CINE acquisition within a single breath-hold of less than 10s and with about 5s reconstruction time. Moreover, CINENet provided visually improved images over CS for high acceleration factors while enabling a 24-fold faster reconstruction. The network generalised for different field of view placements between subjects, SNR levels and mild slice resolution changes.

Demirel et al. (127) proposed applying the Zero-Shot Self-Supervised Learning (ZS-SSL) approach to improve the reconstruction of 8-fold accelerated Cartesian CINE MRI in real-time. ZS-SSL was originally proposed by Yaman et al. (113) to perform self-supervised subject-specific training of a physics-guided reconstruction network without any external training database. This method divides the available measurements from a single scan into three disjoint partitions: two of them are used to enforce data consistency and define the loss while training, while the third is used for self-validation, by defining when to stop training. ZS-SSL significantly outperformed database-trained methods in terms of artefact reduction and generalizability when the training and testing data differed in terms of image characteristics and acquisition parameters.

Yiasemis et al. (114) formulated the CINE reconstruction problem as a least squares regularised optimisation task, and employed the Variable Splitting Half-quadratic ADMM algorithm for Reconstruction of inverse-Problems (vSHARP) to solve it. vSHARP is a DL-based inverse problem solver that incorporates half-quadratic variable splitting and ADMM with neural networks. It uses UNets to replace the need for manually selecting the regularisation prior and learns the solution through denoising steps. Data consistency was enforced with an unrolled gradient descent scheme. This method is optimised in both the image and k-space domains, allowing for high reconstruction fidelity, and achieving an acceleration factor of 10-fold on retrospectively Cartesian undersampled data.

Liu et al. (108) developed the Score-based Self-supervised Diffusion Model (SSJDM) to accelerate 3D CMR acquisitions without fully sampled training data. First, they mapped the undersampled k-space measurements, following a Poisson-disc undersampling scheme, to the corresponding reconstructed images with a self-supervised Bayesian reconstruction network. Secondly, they implemented a joint score-based diffusion model to reconstruct the CMR images, with conditioned Langenvin Markov chain Monte Carlo sampling. The diffusion model gradually injects Gaussian noise into the data, so that it acquires a tractable distribution (i.e. isotropic Gaussian distribution), and is trained to estimate the gradient of this distribution. Then, the reconstruction is accomplished by sampling from the data distribution conditioned on the acquired k-space measurements. After reconstructing the images, dictionary matching was used to obtain $T1$ and $T1\rho$ quantitative maps of the heart. The SSJDM outperformed CS and achieved high quality $T1$ and $T1\rho$ parametric maps close to the reference maps obtained by traditional mapping sequences, even at a high acceleration rate of 14×.

### 7.3.2. Motion Correction

An essential component within any pCMR quantification pipeline involves a motion correction step, which is especially challenging in this context due to image intensity variations caused by the passage of the contrast bolus and low SNR. Classical image registration methods require an optimisation procedure, which implies high computational cost, resulting in a potential task to be leveraged by AI. Recently, several DL-based approaches have been presented to perform

Quantitative First-Pass Perfusion CMR, Feb 2025

rigid and non-rigid image registrations of medical images. However, the field of pCMR has been scarcely explored. Nevertheless, several motion correction methods have been proposed for other CMR modalities, such as CINE (128–131), coronary magnetic resonance angiography (CMRA) (132) and dynamic contrast enhanced (DCE) MR perfusion imaging of other organs/body regions, such as the abdomen (133), liver (134), kidney (135), and breast (136,137).

The image registration problem is usually formulated in image space. Balakrishnan et al. (138) developed VoxelMorph, one of the pioneering works for medical image registration with DL. VoxelMorph is a framework based on the well-known UNet architecture for non-rigid, pairwise image registration. The authors used their method to register 3D MR brain images, but it can be applied to other registration tasks. Morales et al. (128) proposed Cardiac Motion Estimation Network (CarMEN) for non-rigid pairwise registration of a dynamic sequence of 2D CINE images. This unsupervised approach is based on an encoder-decoder CNN, which takes stacks of dynamic multi-slice 2D MR images to estimate 3D cardiac motion fields. Also, Martín-González et al. (129) proposed a DL method named dGW for non-rigid, cardiac motion estimation of a dynamic sequence of 2D CINE images but, in this case, based on a groupwise registration approach. The groupwise approach consists in solving a joint problem to obtain both the motion fields (i.e., transformations of each image in the sequence to the reference) and the reference itself. The architecture of the dGW network is based on a simplified UNet, trained in an unsupervised manner. The loss function is computed by measuring the self-similarity within the sequence with respect to the reference. This self-similarity can be calculated using metrics such as the sum of squared differences or normalised cross-correlation, with the latter being particularly promising for pCMR applications due to its robustness against intensity variations. Qi et al. (139) proposed RespME-net, an unsupervised DL-based method for estimating 3D non-rigid respiratory motion fields with the goal of obtaining motion-corrected CMRA images from free-breathing acquisitions. The proposed network is based on an encoder-decoder CNN architecture, which processes the data as 3D patches extracted from CMRA volumes and outputs the motion fields between the corresponding patches.

An alternative option consists in directly deriving motion fields from k-space data, which could be especially interesting when high acceleration factors are expected, since initially reconstructed images present poor quality. Küstner et al. (140) have recently proposed LAPNet, a non-rigid, pairwise registration method, which exploits the idea that a global non-rigid deformation can be considered as local translational displacements combined with the shift property of the Fourier transform (i.e., a translation in image space corresponds to a linear phase shift in the k-space) in order to formulate the registration in k-space. The architecture of the network includes several convolutional layers, and the last layer performs a fully connected regression to estimate the in-plane deformations at the central location of the k-space input patch. The method was applied to estimate respiratory motion in 3D abdominal MR images.

A completely different approach consists in removing motion artefacts in the motion-corrupted images without the need for the estimation of motion fields. Thus, some DL methods (131,141,142) take motion-corrupted images as input and compute the desired motion-free images as output. To train these methods, the motion-free images can be synthetically corrupted by simulating motion artefacts in k-space. Therefore, the motion-free images can be used as ground-truth for supervised learning.

### Motion-compensated reconstruction

Motion-compensated reconstructions often take advantage of the periodicity of physiological motion to improve the quality of the reconstructed images. These methods typically include both reconstruction and registration steps. In some cases, the methods alternate between both steps, so that a synergistic performance is established in which the reconstruction quality is improved with more accurate motion estimation (143,144).

Qi et al. (132) proposed a motion-corrected DL reconstruction framework for 3D free--breathing CMRA (MoCo-MoDL), consisting of a non-rigid registration network based on the aforementioned RespME-net and a motion--informed model-based reconstruction network based on MoDL, trained in a supervised manner. Huang et al. (145) proposed a DL framework (MODRN) composed of three subnetworks, namely, Dynamic Reconstruction Network (DRN), motion estimation, and motion compensation. The first one consists of an RNN, which takes as input the zero-filled images and outputs initial reconstructions, as an alternative to traditional CS-based algorithms. Next, the motion estimation network takes as input a pair of frames to learn the non-rigid motion between both. Last, the MC network generates motion-guided refined images. In the improved version MODRN(e2e), these components are combined and trained end-to-end. The authors applied their method to 2D CINE images.

Yang et al. (146) and Pan et al. (130) proposed respective unrolled DL frameworks. In (146) each iteration consists of a groupwise non-rigid registration network and a motion-informed reconstruction network, which are trained end-to-end. The registration network is optimised to estimate the invertible motion fields between the whole 2D CINE sequence and an implicit template. Thus, a new dynamic sequence is created by applying these estimated motion fields to the template. This new sequence is incorporated into the reconstruction network to improve the reconstruction quality. In (130), a non-rigid groupwise cardiac motion estimation network is unrolled within an iterative optimisation procedure for 2D CINE reconstruction.

### 7.3.3. Estimation of the Concentration Function

Van Herten et al. (125) developed an open-source physics-informed network to estimate the contrast agent concentration from reconstructed cardiac perfusion images, and then obtain perfusion maps using a two-compartment model. The network consisted of fully connected layers, all followed by a hyperbolic tangent activation function and a batch normalisation layer. The model estimates the contrast agent concentration in the plasma and interstitial space, and the arterial input function at each time point, compares them with the observed concentrations, and computes the residuals of the differential equations of the two-compartment model. The estimated concentrations can then be fit to the model to infer the plasma flow, fractional plasma volume, fractional interstitial volume, and permeability-surface area product.

### 7.3.4. Image Segmentation

Manual-based segmentation can be very time-consuming, tedious and dependent on the level of expertise. DL-based methods enable automatic segmentation of heart regions, facilitating faster, more accurate, and reproducible analysis workflows.

Xue et al. (119) modified the UNet architecture to detect the left ventricle for correct estimation of the arterial input function. The network consisted of a series of residual blocks in downsampling and upsampling branches, connected by skip connections, that received a time series of AIF images and returned the segmentation as output. They also applied this architecture to detect the left ventricle, the myocardium, and the right ventricle through segmentation of a time series of motion-corrected stress and rest perfusion images (120).

Jacobs et al. (121) developed an automated pipeline for segmenting the left ventricle myocardium from myocardial blood flow maps. The right ventricle insertion was first located via landmark detection. Then, the myocardium of perfusion images was segmented with a region-growing algorithm, edge detection, PCA, and active contours. Finally, the segmentation was extended from the perfusion images to the myocardial blood flow maps.

Yalcinkaya et al. (122) proposed the patch3-UNet, a UNet variation training with spatiotemporal (2D + time) patches to segment motion-corrected perfusion images into left ventricle, right ventricle, myocardium, and background. In patch-level training, patches are extracted from the input images and fed to the network; during testing, the network segments the input into patches, outputs the 2D segmentation result for each patch, and combines the segmentations back together to yield the final segmentation result. The proposed strategy also generates an image-based uncertainty map that allows the user to detect the instances where the test data is not suitable for the trained network. The network included an intensity modulation data augmentation strategy to improve generalisation: images were modulated with intensity maps designed to include patterns similar to lateral-wall signal drop-off often seen in raw perfusion images. It was shown that performing segmentation with spatiotemporal patches performs superiorly to the 2D approach since the former learns the temporal connections inherent in the perfusion image series.

Kim et al. (123) modified the UNet architecture to segment the myocardium of dynamic contrast-enhanced myocardial perfusion data, in order to later calculate the myocardial perfusion reserve index. In a first step, the detection of the right ventricle enhanced frame was accomplished with k-means clustering. Then, the myocardium was segmented with a UNet including a Monte Carlo dropout, to obtain the left ventricle centre point, and finally the right ventricle insertion point was calculated through random forest classifiers to allow splitting the myocardium into six uniform segments.

García-Jara et al. (124) proposed to train a UNet with CINE images to perform myocardial segmentation and use this network as a starting point for fine-tuning a second model applied to perfusion images. Through transfer learning, this strategy leverages the wealth of information available from large, publicly accessible CINE datasets, which provide images anatomically analogous to perfusion ones, alleviating the need for large amounts of labelled perfusion data.

Fine-tuning achieved high quality segmentations of the endocardium and epicardium with only 289 perfusion training images available.

### 7.3.5. Clinical Decision Making

DL methods can also be applied to aid diagnosis and clinical decisions. Alskaf et al. (147) proposed applying a hybrid neural network (HNN) to combine image pixel data and clinical information to predict CAD. The HNN combined a CNN to extract image features with a multilayer perceptron to extract clinical data. The HNN prediction was superior to using clinical data alone, indicating DL methods can be used as an auxiliary tool in the diagnostic process.

# 8. Automated Workflows

As new techniques are developed to improve the current pCMR protocols, implementing efficient systematic clinical pipelines becomes increasingly important to save scanner time and reduce the inter-observer variability (148). In fact, the processing pipeline of quantitative pCMR data requires several complex and time-consuming steps that are user-dependent. Many attempts have been made to automate each step of the pipeline, namely the motion correction, AIF correction, left ventricle blood pool segmentation, estimation of the myocardial flow, and segmentation of the myocardium and definition of the segmental distribution of the MBF according to the American Heart Association (AHA) 16-segments model (149). This can be especially useful when working with pixel-wise perfusion mapping.

Automated workflows try to incorporate multiple steps in an automatic form. One first step in that direction was the implementation of reconstruction, motion correction, segmentation of the left ventricle blood pool and calculation of myocardial blood flow within the Gadgetron framework, which allowed for a free-breathing acquisition with a fully automatic pipeline (50).

Later efforts by Hsu et al. (150) integrated a vendor-independent, fully automated and quantitative processing of rest/stress pCMR, including a well validated non-rigid motion correction technique (151) that can be used with different sequences, AIF and myocardial region of interest detection, and the deconvolution of the AIF and the myocardial time-signal curves to calculate the MBF maps. The framework was tested on 17 healthy subjects and 80 CAD patients with good agreement between the automatic and manual approach. In this work, no T2* correction was performed, nor was the time-signal intensity curve directly converted to concentration units, since the MBF can be accurately derived based on mean transit time. A later study with a similar fully automatic pixel-wise approach concluded that quantitative pCMR was able to identify hemodynamically significant CAD when compared with the measured fractional flow reserve (152).

More recently, Xue et al. (153) implemented a pipeline to perform T2* loss correction, motion correction, AIF detection and pixel-wise mapping inline in a clinical scanner, in a fully automated way, from free-breathing dual-sequence acquisitions. This framework was later extended (119) to include a UNet to independently detect the left ventricle in low-resolution AIF images with an accuracy of 99.98% (tested in 1500 healthy subjects), integrated into MR scanners using the Gadgetron Inline AI toolbox (153). Then, Kotecha et al. used this pipeline to demonstrate that quantitative myocardial perfusion mapping increased the accuracy of CMR in differentiating between three- or two-vessel disease from single-vessel CAD (154). Meanwhile, Scannell et al. (155) implemented a DL-based pipeline to detect the left ventricle cavity and myocardium regions, perform motion correction, segment the motion-corrected images, detect the right ventricle insertion points and extract the AIF automatically. This framework would later be modified (55) to include a Dixon reconstruction technique that allowed the acquisition of fat images used to estimate rigid-body respiratory motion to correct for motion in the water images, which performed similarly to spatiotemporal-based registration but was faster to execute.

Jacobs et al. (121) developed a pipeline to perform non-rigid body image registration for motion correction, automatic AIF detection and right ventricle insertion, myocardium segmentation in perfusion images, estimation of myocardial blood flow maps and segmentation of those maps into AHA sectors for analysis. The segmentation found for the myocardial images was used as an input to refine the segmentation of the maps and identify landmarks that allow a segmental analysis (for the AHA segmental model). Tourais et al. (56) proposed to include low-rank and sparsity constrained reconstruction with motion correction into a fully automated inline pipeline to enable free-breathing and high-resolution quantitative pCMR.

Recently, Crawley et al. (56) investigated the use of a fully automated framework to combine automated quantitative pCMR mapping with high-resolution free-breathing perfusion imaging, generated in-line on the scanner, in order to identify functionally significant CAD. Image reconstruction was performed iteratively with motion correction, while a CNN was employed to identify the left ventricle, a bounding box to crop the myocardial and AIF image series, and the blood pool was segmented automatically.

The use of an automated pipeline that does not depend on the operator training level is key to improve the reproducibility of quantitative pCMR. In fact, Brown et al. showed that when using fully automatic methodology, the reproducibility is similar to the one presented in PET, with MBF showing better repeatability than myocardial perfusion reserve indicators (156). They applied a pipeline that included reconstruction, motion correction, AIF segmentation, conversion of signal into contrast concentration and perfusion quantification, which took less than 3 minutes; however, the subsequent analysis that included myocardial segmentation and segment definition was performed manually offline.

Quantitative First-Pass Perfusion CMR, Feb 2025

Lastly, the usefulness of using a high-resolution AHA model that allows the evaluation of the intra-segment distribution of the MBF was demonstrated by Scannell et al. to be useful for a detailed identification of perfusion defects (157). Their work also used a fully automatic high resolution, quantitative stress myocardial methodology previously used for paediatric patients.

# 9. Conclusions

CMR has been demonstrated to be a comprehensive exam that provides evaluation of cardiovascular anatomy, function, tissue composition, and blood flow. Compared to alternative techniques, CMR can assess myocardial viability in patients with CAD and identify cases where there is infarction with non-obstructive coronary arteries, without the need for ionizing radiation. Furthermore, pCMR has the potential of offering a quantitative analysis to assess myocardial blood flow in an observer-independent manner, showing promise to diagnose both epicardial and microvascular CAD in an observer-independent way. This review has provided an overview of the main quantitative pCMR concepts and techniques, from data acquisition to post-processing tools. In addition, we discussed several trends in the workflow, including automated motion-corrected image reconstruction and DL-based reconstruction for improved resolution and heart coverage.

pCMR is performed by imaging the heart during the first-pass of a contrast agent, with a T1-sensitive acquisition sequence. Unlike non-quantitative methods, not only is it necessary to measure the contrast reaching the myocardial tissue, but an accurate measurement of the contrast distributed to the heart (i.e. AIF) is also required. Two main acquisition sequences have been proposed for this purpose: the dual-bolus and the dual-saturation sequences. Moreover, the presence of partial volume effects and dark rim artefacts must be accounted for when designing sequences and in post-processing in order to avoid misinterpretation of perfusion defects.

Acquiring a quantitative map for pCMR can be accomplished with a large collection of tracer-kinetic models, each with advantages and limitations (e.g. inability to estimate plasma flow and tissue permeability separately, complex calculations, time and software demands). While choosing a gold-standard tracer-kinetic model is no trivial task, integrating pCMR into the clinical practice will require more standardised result-reporting across protocols.

Arguably one of the most challenging limitations of pCMR is its limited spatial and temporal resolutions and heart coverage. Multiple techniques have been proposed to tackle these challenges, most notably methods that allow reconstructing signal-intensity images from a smaller amount of acquired data, so that the accrued extra time can be used to obtain more information on cardiovascular anatomy or physiology. Amongst the proposed methods, state-of-the-art techniques seem to lean into a combination of several strategies that include parallel imaging and CS or LR, as well as incorporating physics knowledge into model-based reconstruction approaches.

Accelerated image reconstruction has also been increasingly combined with motion-correction, potentially enabling free-breathing acquisitions, which are not constrained by the patient's ability to sustain their breath and would greatly increase the current heart coverage limit.

Future directions in reconstruction and motion correction also seem to point to adopting AI strategies. While methods developed specifically for pCMR are still limited and mostly based on simpler network architectures, there is a growing body of work developed for CMR in general which can potentially be adapted for pCMR in the future. However, these novel methods still require further testing in multi-centre, multi-vendor conditions before their integration into clinical routine.

Finally, pCMR involves a complex protocol where multiple choices must be made along the way regarding data acquisition, acceleration, and quantification. In the interest of standardising and facilitating the workflow, as well as offering a more observer-independent approach, multiple projects have been proposed to automate the pCMR pipeline, either through generating quantitative perfusion maps in-line or through fast and easy-to-use analysis tools. These automated pipelines often include accelerated image reconstruction, motion correction, AIF estimation, segmentation of regions of interest and pixel-wise calculation of myocardial blood flow, automatically accelerating tasks that would otherwise be cumbersome for clinicians and potentially allowing the standardisation of pCMR protocols.

# Funding Sources

This Project received funding from 'la Caixa' Foundation and FCT, I P under the Project code LCF/PR/HR22/00533, Portuguese national funds from FCT - Foundation for Science and Technology through projects UIDB/04326/2020 (DOI:10.54499/UIDB/04326/2020), UIDP/04326/2020 (DOI:10.54499/UIDP/04326/2020) and LA/P/0101/2020 (DOI:10.54499/LA/P/0101/2020), and LARSyS FCT funding (DOI: 10.54499/LA/P/0083/2020, 10.54499/UIDP/50009/2020, and 10.54499/UIDB/50009/2020). This work was supported in part by the Agencia Estatal de Investigacion, under Grants PID2020-115339RB-I00 and TED2021-130090B-I00, and by Fundacion La Caixa (HR22-00533). R.-M. M.-L. is supported by the Spanish MICIU-AEI "Ramón y Cajal" program under grant RYC2023-045078-I.

# References

1. Schwitter J, Wacker CM, Van Rossum AC, Lombardi M, Al-Saadi N, Ahlstrom H, et al. MR-IMPACT: comparison of perfusion-cardiac magnetic resonance with single-photon emission computed tomography for the detection of coronary artery disease in a multicentre, multivendor, randomized trial. Eur Heart J. 2008 Feb;29(4):480–9.

2. Jaarsma C, Leiner T, Bekkers SC, Crijns HJ, Wildberger JE, Nagel E, et al. Diagnostic Performance of Noninvasive Myocardial Perfusion Imaging Using Single-Photon Emission Computed Tomography, Cardiac Magnetic Resonance, and Positron Emission Tomography Imaging for the Detection of Obstructive Coronary Artery Disease. J Am Coll Cardiol. 2012 May;59(19):1719–28.

3. Morton G, Chiribiri A, Ishida M, Hussain ST, Schuster A, Indermuehle A, et al. Quantification of Absolute Myocardial Perfusion in Patients With Coronary Artery Disease. J Am Coll Cardiol. 2012 Oct;60(16):1546–55.

4. Greenwood JP, Maredia N, Younger JF, Brown JM, Nixon J, Everett CC, et al. Cardiovascular magnetic resonance and single-photon emission computed tomography for diagnosis of coronary heart disease (CE-MARC): a prospective trial. The Lancet. 2012 Feb;379(9814):453–60.

5. Pontone G, Guaricci AI, Palmer SC, Andreini D, Verdecchia M, Fusini L, et al. Diagnostic performance of non-invasive imaging for stable coronary artery disease: A meta-analysis. Int J Cardiol. 2020 Feb;300:276–81.

6. Rajiah PS, François CJ, Leiner T. Cardiac MRI: State of the Art. Radiology. 2023 May 1;307(3):e223008.

7. Jerosch-Herold M. Quantification of myocardial perfusion by cardiovascular magnetic resonance. J Cardiovasc Magn Reson. 2010 Dec;12(1):57.

8. Fair MJ, Gatehouse PD, DiBella EVR, Firmin DN. A review of 3D first-pass, whole-heart, myocardial perfusion cardiovascular magnetic resonance. J Cardiovasc Magn Reson. 2015 Dec;17(1):68.

9. Heydari B, Kwong RY, Jerosch-Herold M. Technical Advances and Clinical Applications of Quantitative Myocardial Blood Flow Imaging With Cardiac MRI. Prog Cardiovasc Dis. 2015 May;57(6):615–22.

10. Patel AR, Salerno M, Kwong RY, Singh A, Heydari B, Kramer CM. Stress Cardiac Magnetic Resonance Myocardial Perfusion Imaging. J Am Coll Cardiol. 2021 Oct;78(16):1655–68.

11. Alskaf E, Dutta U, Scannell CM, Chiribiri A. Deep learning applications in myocardial perfusion imaging, a systematic review and meta-analysis. Inform Med Unlocked. 2022;32:101055.

12. Pons-Lladó G, Kellman P. State-of-the-Art of Myocardial Perfusion by CMR: A Practical View. Rev Cardiovasc Med. 2022 Sep 26;23(10):325.

13. Sharrack N, Chiribiri A, Schwitter J, Plein S. How to do quantitative myocardial perfusion cardiovascular magnetic resonance. Eur Heart J - Cardiovasc Imaging. 2022 Feb 22;23(3):315–8.

14. Del Buono MG, Montone RA, Camilli M, Carbone S, Narula J, Lavie CJ, et al. Coronary Microvascular Dysfunction Across the Spectrum of Cardiovascular Diseases. J Am Coll Cardiol. 2021 Sep;78(13):1352–71.

15. Kotecha T, Martinez-Naharro A, Boldrini M, Knight D, Hawkins P, Kalra S, et al. Automated Pixel-Wise Quantitative Myocardial Perfusion Mapping by CMR to Detect Obstructive Coronary Artery Disease and Coronary Microvascular Dysfunction. JACC Cardiovasc Imaging. 2019 Oct;12(10):1958–69.

16. Ibanez B, Aletras AH, Arai AE, Arheden H, Bax J, Berry C, et al. Cardiac MRI Endpoints in Myocardial Infarction Experimental and Clinical Trials. J Am Coll Cardiol. 2019 Jul;74(2):238–56.

17. Hundley WG, Bluemke DA, Finn JP, Flamm SD, Fogel MA, Friedrich MG, et al. ACCF/ACR/AHA/NASCI/SCMR 2010 Expert Consensus Document on Cardiovascular Magnetic Resonance. J Am Coll Cardiol. 2010 Jun;55(23):2614–62.

18. Arai AE, Schulz-Menger J, Shah DJ, Han Y, Bandettini WP, Abraham A, et al. Stress Perfusion Cardiac Magnetic Resonance vs SPECT Imaging for Detection of Coronary Artery Disease. J Am Coll Cardiol. 2023 Nov;82(19):1828–38.

19. Sánchez-González J, Fernandez-Jiménez R, Nothnagel ND, López-Martín G, Fuster V, Ibañez B. Optimization of dual-saturation single bolus acquisition for quantitative cardiac perfusion and myocardial blood flow maps. J Cardiovasc Magn Reson. 2015 Jan;17(1):21.

20. Kotecha T, Monteagudo JM, Martinez-Naharro A, Chacko L, Brown J, Knight D, et al. Quantitative cardiovascular magnetic resonance myocardial perfusion mapping to assess hyperaemic response to adenosine stress. Eur Heart J - Cardiovasc Imaging. 2021 Feb 22;22(3):273–81.

21. Brown LAE, Gulsin GS, Onciul SC, Broadbent DA, Yeo JL, Wood AL, et al. Sex- and age-specific normal values for automated quantitative pixel-wise myocardial perfusion cardiovascular magnetic resonance. Eur Heart J - Cardiovasc Imaging. 2023 Mar 21;24(4):426–34.

22. Gulati M, Levy PD, Mukherjee D, Amsterdam E, Bhatt DL, Birtcher KK, et al. 2021 AHA/ACC/ASE/CHEST/SAEM/SCCT/SCMR Guideline for the Evaluation and Diagnosis of Chest Pain. J Am Coll Cardiol. 2021 Nov;78(22):e187–285.

23. Knuuti J. 2019 ESC Guidelines for the diagnosis and management of chronic coronary syndromes The Task Force for the diagnosis and management of chronic coronary syndromes of the European Society of Cardiology (ESC). Russ J Cardiol. 2020 Mar 11;25(2):119–80.

24. Pasupathy S, Air T, Dreyer RP, Tavella R, Beltrame JF. Systematic Review of Patients Presenting With Suspected Myocardial Infarction and Nonobstructive Coronary Arteries. Circulation. 2015 Mar 10;131(10):861–70.

25. Ferreira VM, Schulz-Menger J, Holmvang G, Kramer CM, Carbone I, Sechtem U, et al. Cardiovascular Magnetic Resonance in Nonischemic Myocardial Inflammation. J Am Coll Cardiol. 2018 Dec;72(24):3158–76.

26. Sara JD, Widmer RJ, Matsuzawa Y, Lennon RJ, Lerman LO, Lerman A. Prevalence of Coronary Microvascular Dysfunction Among Patients With Chest Pain and Nonobstructive Coronary Artery Disease. JACC Cardiovasc Interv. 2015 Sep;8(11):1445–53.

27. Camici PG, Tschöpe C, Di Carli MF, Rimoldi O, Van Linthout S. Coronary microvascular dysfunction in hypertrophy and heart failure. Cardiovasc Res. 2020 Mar 1;116(4):806–16.

28. Shah SJ, Lam CSP, Svedlund S, Saraste A, Hage C, Tan RS, et al. Prevalence and correlates of coronary microvascular dysfunction in heart failure with preserved ejection fraction: PROMIS-HFpEF. Eur Heart J. 2018 Oct 1;39(37):3439–50.

29. Lelovas PP, Kostomitsopoulos NG, Xanthos TT. A Comparative Anatomic and Physiologic Overview of the Porcine Heart. J Am Assoc Lab Anim Sci. 2014;53(5).

30. Fernández-Jiménez R, Silva J, Martínez-Martínez S, López-Maderuelo MD, Nuno-Ayala M, García-Ruiz JM, et al. Impact of Left Ventricular Hypertrophy on Troponin Release During Acute Myocardial Infarction: New Insights From a Comprehensive Translational Study. J Am Heart Assoc. 2015 Jan 5;4(1):e001218.

31. Galán-Arriola C, Lobo M, Vílchez-Tschischke JP, López GJ, De Molina-Iracheta A, Pérez-

Martínez C, et al. Serial Magnetic Resonance Imaging to Identify Early Stages of Anthracycline-Induced Cardiotoxicity. J Am Coll Cardiol. 2019 Feb;73(7):779–91.

32. Quintanilla JG, Alfonso-Almazán JM, Pérez-Castellano N, Pandit SV, Jalife J, Pérez-Villacastín J, et al. Instantaneous Amplitude and Frequency Modulations Detect the Footprint of Rotational Activity and Reveal Stable Driver Regions as Targets for Persistent Atrial Fibrillation Ablation. Circ Res. 2019 Aug 30;125(6):609–27.

33. Martínez-Milla J, Galán-Arriola C, Carnero M, Cobiella J, Pérez-Camargo D, Bautista-Hernández V, et al. Translational large animal model of hibernating myocardium: characterization by serial multimodal imaging. Basic Res Cardiol. 2020 May;115(3):33.

34. Porta-Sánchez A, Mazzanti A, Tarifa C, Kukavica D, Trancuccio A, Mohsin M, et al. Unexpected impairment of INa underpins reentrant arrhythmias in a knock-in swine model of Timothy syndrome. Nat Cardiovasc Res. 2023 Dec 11;2(12):1291–309.

35. Varga-Szemes A, Kiss P, Brott BC, Wang D, Simor T, Elgavish GA. Embozene™ microspheres induced nonreperfused myocardial infarction in an experimental swine model. Catheter Cardiovasc Interv. 2013 Mar;81(4):689–97.

36. Galán-Arriola C, Vílchez-Tschischke JP, Lobo M, López GJ, De Molina-Iracheta A, Pérez-Martínez C, et al. Coronary microcirculation damage in anthracycline cardiotoxicity. Cardiovasc Res. 2022 Jan 29;118(2):531–41.

37. Renner S, Martins AS, Streckel E, Braun-Reichhart C, Backman M, Prehn C, et al. Mild maternal hyperglycemia in *INS* C93S transgenic pigs causes impaired glucose tolerance and metabolic alterations in neonatal offspring. Dis Model Mech. 2019 Jan 1;dmm.039156.

38. Dorado B, Pløen GG, Barettino A, Macías A, Gonzalo P, Andrés-Manzano MJ, et al. Generation and characterization of a novel knockin minipig model of Hutchinson-Gilford progeria syndrome. Cell Discov. 2019 Mar 19;5(1):16.

39. Ibanez B, Aletras AH, Arai AE, Arheden H, Bax J, Berry C, et al. Cardiac MRI Endpoints in Myocardial Infarction Experimental and Clinical Trials. J Am Coll Cardiol. 2019 Jul;74(2):238–56.

40. Bobi J, Solanes N, Fernández-Jiménez R, Galán-Arriola C, Dantas AP, Fernández-Friera L, et al. Intracoronary Administration of Allogeneic Adipose Tissue–Derived Mesenchymal Stem Cells Improves Myocardial Perfusion But Not Left Ventricle Function, in a Translational Model of Acute Myocardial Infarction. J Am Heart Assoc. 2017 May 5;6(5):e005771.

41. Alvino VV, Fernández-Jiménez R, Rodriguez-Arabaolaza I, Slater S, Mangialardi G, Avolio E, et al. Transplantation of Allogeneic Pericytes Improves Myocardial Vascularization and Reduces Interstitial Fibrosis in a Swine Model of Reperfused Acute Myocardial Infarction. J Am Heart Assoc. 2018 Jan 23;7(2):e006727.

42. Jin H, Yun H, Ma J, Chen Z, Chang S, Zeng M. Coronary Microembolization with Normal

Epicardial Coronary Arteries and No Visible Infarcts on Nitrobluetetrazolium Chloride-Stained Specimens: Evaluation with Cardiac Magnetic Resonance Imaging in a Swine Model. Korean J Radiol. 2016;17(1):83.

43. Van Houten M, Yang Y, Hauser A, Glover DK, Gan L, Yeager M, et al. Adenosine stress CMR perfusion imaging of the temporal evolution of perfusion defects in a porcine model of progressive obstructive coronary artery occlusion. NMR Biomed. 2019 Nov;32(11):e4136.

44. Do HP, Ramanan V, Qi X, Barry J, Wright GA, Ghugre NR, et al. Non-contrast assessment of microvascular integrity using arterial spin labeled cardiovascular magnetic resonance in a porcine model of acute myocardial infarction. J Cardiovasc Magn Reson. 2018 Feb;20(1):45.

45. Jerosch-Herold M. Quantification of myocardial perfusion by cardiovascular magnetic resonance. J Cardiovasc Magn Reson. 2010 Oct;12(1):57.

46. Kellman P, Arai AE. Imaging Sequences for First Pass Perfusion - A Review. J Cardiovasc Magn Reson. 2007 May;9(3):525–37.

47. Milidonis X, Franks R, Schneider T, Sánchez-González J, Sammut EC, Plein S, et al. Influence of the arterial input sampling location on the diagnostic accuracy of cardiovascular magnetic resonance stress myocardial perfusion quantification. J Cardiovasc Magn Reson. 2021 Mar;23(1):35.

48. Gatehouse PD, Elkington AG, Ablitt NA, Yang G, Pennell DJ, Firmin DN. Accurate assessment of the arterial input function during high-dose myocardial perfusion cardiovascular magnetic resonance. J Magn Reson Imaging. 2004 Jul;20(1):39–45.

49. Hsu L, Rhoads KL, Holly JE, Kellman P, Aletras AH, Arai AE. Quantitative myocardial perfusion analysis with a dual-bolus contrast-enhanced first-pass MRI technique in humans. J Magn Reson Imaging. 2006 Mar;23(3):315–22.

50. Kellman P, Hansen MS, Nielles-Vallespin S, Nickander J, Themudo R, Ugander M, et al. Myocardial perfusion cardiovascular magnetic resonance: optimized dual sequence and reconstruction for quantification. J Cardiovasc Magn Reson. 2016 Dec;19(1):43.

51. Kellman P, Aletras AH, Hsu L, McVeigh ER, Arai AE. $T$ measurement during first-pass contrast-enhanced cardiac perfusion imaging. Magn Reson Med. 2006 Nov;56(5):1132–4.

52. Nielles-Vallespin S, Kellman P, Hsu LY, Arai AE. FLASH proton density imaging for improved surface coil intensity correction in quantitative and semi-quantitative SSFP perfusion cardiovascular magnetic resonance. J Cardiovasc Magn Reson. 2015 Jan;17(1):16.

53. Cernicanu A, Axel L. Theory-Based Signal Calibration with Single-Point T1 Measurements for First-Pass Quantitative Perfusion MRI Studies. Acad Radiol. 2006 Jun;13(6):686–93.

54. Fair MJ, Gatehouse PD, DiBella EVR, Firmin DN. A review of 3D first-pass, whole-heart, myocardial perfusion cardiovascular magnetic resonance. J Cardiovasc Magn Reson. 2015 Jan;17(1):68.

55. Scannell CM, Correia T, Villa ADM, Schneider T, Lee J, Breeuwer M, et al. Feasibility of free-breathing quantitative myocardial perfusion using multi-echo Dixon magnetic resonance imaging. Sci Rep. 2020 Jul 29;10(1):12684.

56. Tourais J, Scannell CM, Schneider T, Alskaf E, Crawley R, Bosio F, et al. High-Resolution Free-Breathing Quantitative First-Pass Perfusion Cardiac MR Using Dual-Echo Dixon With Spatio-Temporal Acceleration. Front Cardiovasc Med. 2022 Apr 29;9:884221.

57. Ta AD, Hsu LY, Conn HM, Winkler S, Greve AM, Shanbhag SM, et al. Fully quantitative pixel-wise analysis of cardiovascular magnetic resonance perfusion improves discrimination of dark rim artifact from perfusion defects associated with epicardial coronary stenosis. J Cardiovasc Magn Reson. 2018 Feb;20(1):16.

58. Ferreira P, Gatehouse P, Kellman P, Bucciarelli-Ducci C, Firmin D. Variability of myocardial perfusion dark rim Gibbs artifacts due to sub-pixel shifts. J Cardiovasc Magn Reson. 2009 Jan;11(1):17.

59. Storey P, Chen Q, Li W, Edelman RR, Prasad PV. Band artifacts due to bulk motion. Magn Reson Med. 2002 Dec;48(6):1028–36.

60. Di Bella EVR, Parker DL, Sinusas AJ. On the dark rim artifact in dynamic contrast-enhanced MRI myocardial perfusion studies. Magn Reson Med. 2005 Nov;54(5):1295–9.

61. Motwani M, Maredia N, Fairbairn TA, Kozerke S, Radjenovic A, Greenwood JP, et al. High-Resolution Versus Standard-Resolution Cardiovascular MR Myocardial Perfusion Imaging for the Detection of Coronary Artery Disease. Circ Cardiovasc Imaging. 2012 May;5(3):306–13.

62. Jerosch-Herold M, Swingen C, Seethamraju RT. Myocardial blood flow quantification with MRI by model-independent deconvolution. Med Phys. 2002 May;29(5):886–97.

63. Larsson HBW, Stubgaard M, Frederiksen JL, Jensen M, Henriksen O, Paulson OB. Quantitation of blood-brain barrier defect by magnetic resonance imaging and gadolinium-DTPA in patients with multiple sclerosis and brain tumors. Magn Reson Med. 1990 Oct;16(1):117–31.

64. Patlak CS, Blasberg RG, Fenstermacher JD. Graphical Evaluation of Blood-to-Brain Transfer Constants from Multiple-Time Uptake Data. J Cereb Blood Flow Metab. 1983 Mar;3(1):1–7.

65. Tofts PS, Kermode AG. Measurement of the blood-brain barrier permeability and leakage space using dynamic MR imaging. 1. Fundamental concepts. Magn Reson Med. 1991 Feb;17(2):357–67.

66. Tofts PS. Modeling tracer kinetics in dynamic Gd-DTPA MR imaging. J Magn Reson Imaging. 1997 Jan;7(1):91–101.

67. Brix G, Semmler W, Port R, Schad LR, Layer G, Lorenz WJ. Pharmacokinetic Parameters

in CNS Gd-DTPA Enhanced MR Imaging: J Comput Assist Tomogr. 1991 Jul;15(4):621–8.

68. Khalifa F, Soliman A, El-Baz A, Abou El-Ghar M, El-Diasty T, Gimel'farb G, et al. Models and methods for analyzing DCE-MRI: A review. Med Phys. 2014 Dec;41(12):124301.

69. Johnson J, Wilson T. A model for capillary exchange. Am J Physiol-Leg Content. 1966 Jun 1;210(6):1299–303.

70. Sangren WC, Sheppard CW. A mathematical derivation of the exchange of a labeled substance between a liquid flowing in a vessel and an external compartment. Bull Math Biophys. 1953 Dec;15(4):387–94.

71. Bassingthwaighte JB, Wang CY, Chan IS. Blood-tissue exchange via transport and transformation by capillary endothelial cells. Circ Res. 1989 Oct;65(4):997–1020.

72. Ismail TF, Strugnell W, Coletti C, Božić-Iven M, Weingärtner S, Hammernik K, et al. Cardiac MR: From Theory to Practice. Front Cardiovasc Med. 2022 Mar 3;9:826283.

73. Pruessmann KP, Weiger M, Scheidegger MB, Boesiger P. SENSE: Sensitivity encoding for fast MRI. Magn Reson Med. 1999 Nov;42(5):952–62.

74. Lustig M, Donoho D, Pauly JM. Sparse MRI: The application of compressed sensing for rapid MR imaging. Magn Reson Med. 2007 Dec;58(6):1182–95.

75. Zhao B, Haldar JP, Brinegar C, Liang ZP. Low rank matrix recovery for real-time cardiac MRI. In: 2010 IEEE International Symposium on Biomedical Imaging: From Nano to Macro [Internet]. Rotterdam, Netherlands: IEEE; 2010 [cited 2024 Aug 30]. p. 996–9. Available from: http://ieeexplore.ieee.org/document/5490156/

76. Lingala SG, Hu Y, DiBella E, Jacob M. Accelerated Dynamic MRI Exploiting Sparsity and Low-Rank Structure: k-t SLR. IEEE Trans Med Imaging. 2011 May;30(5):1042–54.

77. Otazo R, Candès E, Sodickson DK. Low-rank plus sparse matrix decomposition for accelerated dynamic MRI with separation of background and dynamic components: L+S Reconstruction. Magn Reson Med. 2015 Mar;73(3):1125–36.

78. Fessler J. Model-Based Image Reconstruction for MRI. IEEE Signal Process Mag. 2010 Jul;27(4):81–9.

79. Pedersen H, Kozerke S, Ringgaard S, Nehrke K, Kim WY. *k-t* PCA: Temporally constrained *k-t* BLAST reconstruction using principal component analysis. Magn Reson Med. 2009 Sep;62(3):706–16.

80. Dikaios N, Arridge S, Hamy V, Punwani S, Atkinson D. Direct parametric reconstruction from undersampled (k, t)-space data in dynamic contrast enhanced MRI. Med Image Anal. 2014 Oct;18(7):989–1001.

81. Guo Y, Lingala SG, Zhu Y, Lebel RM, Nayak KS. Direct estimation of tracer-kinetic

parameter maps from highly undersampled brain dynamic contrast enhanced MRI. Magn Reson Med. 2017 Oct;78(4):1566–78.

82. Correia T, Schneider T, Chiribiri A. Model-Based Reconstruction for Highly Accelerated First-Pass Perfusion Cardiac MRI. In: Shen D, Liu T, Peters TM, Staib LH, Essert C, Zhou S, et al., editors. Medical Image Computing and Computer Assisted Intervention – MICCAI 2019 [Internet]. Cham: Springer International Publishing; 2019 [cited 2023 Jun 18]. p. 514–22. (Lecture Notes in Computer Science; vol. 11765). Available from: https://link.springer.com/10.1007/978-3-030-32245-8_57

83. Shin T, Nayak KS, Santos JM, Nishimura DG, Hu BS, McConnell MV. Three-dimensional first-pass myocardial perfusion MRI using a stack-of-spirals acquisition. Magn Reson Med. 2013 Mar;69(3):839–44.

84. Vitanis V, Manka R, Giese D, Pedersen H, Plein S, Boesiger P, et al. High resolution three-dimensional cardiac perfusion imaging using compartment-based *k-t* principal component analysis. Magn Reson Med. 2011 Feb;65(2):575–87.

85. Schmidt JFM, Wissmann L, Manka R, Kozerke S. Iterative k-t principal component analysis with nonrigid motion correction for dynamic three-dimensional cardiac perfusion imaging: Iterative k-t PCA with Nonrigid Motion Correction. Magn Reson Med. 2014 Jul;72(1):68–79.

86. Chen L, Adluru G, Schabel MC, McGann CJ, DiBella EVR. Myocardial perfusion MRI with an undersampled 3D stack-of-stars sequence. Med Phys. 2012 Aug;39(8):5204–11.

87. Stäb D, Wech T, Breuer FA, Weng AM, Ritter CO, Hahn D, et al. High resolution myocardial first-pass perfusion imaging with extended anatomic coverage: High Coverage Myocardial Perfusion MRI. J Magn Reson Imaging. 2014 Jun;39(6):1575–87.

88. McElroy S, Ferrazzi G, Nazir MS, Kunze KP, Neji R, Speier P, et al. Combined simultaneous multislice bSSFP and compressed sensing for first-pass myocardial perfusion at 1.5 T with high spatial resolution and coverage. Magn Reson Med. 2020 Dec;84(6):3103–16.

89. Chen X, Salerno M, Yang Y, Epstein FH. Motion-compensated compressed sensing for dynamic contrast-enhanced MRI using regional spatiotemporal sparsity and region tracking: Block low-rank sparsity with motion-guidance (BLOSM): BLOSM: Block Low-rank Sparsity with Motion-guidance. Magn Reson Med. 2014 Oct;72(4):1028–38.

90. Tian Y, Mendes J, Pedgaonkar A, Ibrahim M, Jensen L, Schroeder JD, et al. Feasibility of multiple-view myocardial perfusion MRI using radial simultaneous multi-slice acquisitions. Tang D, editor. PLOS ONE. 2019 Feb 11;14(2):e0211738.

91. Akçakaya M, Basha TA, Pflugi S, Foppa M, Kissinger KV, Hauser TH, et al. Localized spatio-temporal constraints for accelerated CMR perfusion. Magn Reson Med. 2014 Sep;72(3):629–39.

92. Christodoulou AG, Shaw JL, Nguyen C, Yang Q, Xie Y, Wang N, et al. Magnetic resonance

multitasking for motion-resolved quantitative cardiovascular imaging. Nat Biomed Eng. 2018 Apr 9;2(4):215–26.

93. Hoh T, Vishnevskiy V, Polacin M, Manka R, Fuetterer M, Kozerke S. Free-breathing motion-informed locally low-rank quantitative 3D myocardial perfusion imaging. Magn Reson Med. 2022 Oct;88(4):1575–91.

94. Sun C, Robinson A, Wang Y, Bilchick KC, Kramer CM, Weller D, et al. A Slice-Low-Rank Plus Sparse ( SLICE-L + S) Reconstruction Method for k-t Undersampled Multiband First-Pass Myocardial Perfusion MRI. Magn Reson Med. 2022 Sep;88(3):1140–55.

95. Shin T, Nayak KS, Santos JM, Nishimura DG, Hu BS, McConnell MV. Three-dimensional first-pass myocardial perfusion MRI using a stack-of-spirals acquisition. Magn Reson Med. 2013 Mar;69(3):839–44.

96. Wang J, Yang Y, Weller DS, Zhou R, Van Houten M, Sun C, et al. High spatial resolution spiral first-pass myocardial perfusion imaging with whole-heart coverage at 3 T. Magn Reson Med. 2021 Aug;86(2):648–62.

97. Wang J, Weller DS, Kramer CM, Salerno M. DEep learning-based rapid Spiral Image REconstruction (DESIRE) for high-resolution spiral first-pass myocardial perfusion imaging. NMR Biomed. 2022 May;35(5):e4661.

98. Harrison A, Adluru G, Damal K, Shaaban AM, Wilson B, Kim D, et al. Rapid ungated myocardial perfusion cardiovascular magnetic resonance: preliminary diagnostic accuracy. J Cardiovasc Magn Reson. 2013 Jan;15(1):26.

99. Sharif B, Arsanjani R, Dharmakumar R, Bairey Merz CN, Berman DS, Li D. All-systolic non-ECG-gated myocardial perfusion MRI: Feasibility of multi-slice continuous first-pass imaging: All-Systolic Non-ECG-gated Myocardial Perfusion MRI. Magn Reson Med. 2015 Dec;74(6):1661–74.

100. Likhite D, Suksaranjit P, Adluru G, Hu N, Weng C, Kholmovski E, et al. Interstudy repeatability of self-gated quantitative myocardial perfusion MRI. J Magn Reson Imaging. 2016 Jun;43(6):1369–78.

101. Tian Y, Mendes J, Wilson B, Ross A, Ranjan R, DiBella E, et al. Whole-heart, ungated, free-breathing, cardiac-phase-resolved myocardial perfusion MRI by using Continuous Radial Interleaved simultaneous Multi-slice acquisitions at sPoiled steady-state (CRIMP). Magn Reson Med. 2020 Dec;84(6):3071–87.

102. Huttinga NRF, Van Den Berg CAT, Luijten PR, Sbrizzi A. MR-MOTUS: model-based non-rigid motion estimation for MR-guided radiotherapy using a reference image and minimal $k$-space data. Phys Med Biol. 2020 Jan 10;65(1):015004.

103. Huttinga NRF, Bruijnen T, Van Den Berg CAT, Sbrizzi A. Nonrigid 3D motion estimation at high temporal resolution from prospectively undersampled k-space data using low-rank MR-MOTUS. Magn Reson Med. 2021 Apr;85(4):2309–26.

104. Huttinga NRF, Bruijnen T, Van Den Berg CAT, Sbrizzi A. Real-Time Non-Rigid 3D Respiratory Motion Estimation for MR-Guided Radiotherapy Using MR-MOTUS. IEEE Trans Med Imaging. 2022 Feb;41(2):332–46.

105. Olausson TE, Terpstra ML, Huttinga NRF, Beijst C, Blanken N, Correia T, et al. Free-Running Time-Resolved First-Pass Myocardial Perfusion Using a Multi-Scale Dynamics Decomposition: CMR-MOTUS.

106. Otazo R, Candès E, Sodickson DK. Low-rank plus sparse matrix decomposition for accelerated dynamic MRI with separation of background and dynamic components: L+S Reconstruction. Magn Reson Med. 2015 Mar;73(3):1125–36.

107. Leiner T, Rueckert D, Suinesiaputra A, Baeßler B, Nezafat R, Išgum I, et al. Machine learning in cardiovascular magnetic resonance: basic concepts and applications. J Cardiovasc Magn Reson. 2019 Jan;21(1):61.

108. Liu Y, Cui ZX, Liu C, Zheng H, Wang H, Zhou Y, et al. Score-based Diffusion Models With Self-supervised Learning For Accelerated 3D Multi-contrast Cardiac Magnetic Resonance Imaging [Internet]. arXiv; 2023 [cited 2023 Nov 27]. Available from: http://arxiv.org/abs/2310.04669

109. Aggarwal HK, Mani MP, Jacob M. MoDL: Model Based Deep Learning Architecture for Inverse Problems. IEEE Trans Med Imaging. 2019 Feb;38(2):394–405.

110. Biswas S, Aggarwal HK, Jacob M. Dynamic MRI using model-based deep learning and SToRM priors: MoDL-SToRM. Magn Reson Med. 2019 Jul;82(1):485–94.

111. Qin C, Schlemper J, Caballero J, Price AN, Hajnal JV, Rueckert D. Convolutional Recurrent Neural Networks for Dynamic MR Image Reconstruction. IEEE Trans Med Imaging. 2019 Jan;38(1):280–90.

112. Küstner T. CINENet: deep learning-based 3D cardiac CINE MRI reconstruction with multi-coil complex-valued 4D spatio-temporal convolutions. Sci Rep. 2020;

113. Yaman B, Hosseini SAH, Akçakaya M. Zero-Shot Self-Supervised Learning for MRI Reconstruction [Internet]. arXiv; 2023 [cited 2023 Dec 13]. Available from: http://arxiv.org/abs/2102.07737

114. Yiasemis G, Moriakov N, Sonke JJ, Teuwen J. Deep Cardiac MRI Reconstruction with ADMM [Internet]. arXiv; 2023 [cited 2023 Nov 23]. Available from: http://arxiv.org/abs/2310.06628

115. Fan L, Shen D, Haji-Valizadeh H, Naresh NK, Carr JC, Freed BH, et al. Rapid dealiasing of undersampled, non-Cartesian cardiac perfusion images using U-net. NMR Biomed. 2020 May;33(5):e4239.

116. Le J, Tian Y, Mendes J, Wilson B, Ibrahim M, DiBella E, et al. Deep learning for radial SMS myocardial perfusion reconstruction using the 3D residual booster U-net. Magn Reson

Imaging. 2021 Nov;83:178–88.

117. Martín-González E, Alskaf E, Chiribiri A, Casaseca-de-la-Higuera P, Alberola-López C, Nunes RG, et al. Physics-informed self-supervised deep learning reconstruction for accelerated first-pass perfusion cardiac MRI. In 2021 [cited 2023 Jun 18]. p. 86–95. Available from: http://arxiv.org/abs/2301.02033

118. Demirel OB, Yaman B, Shenoy C, Moeller S, Weingärtner S, Akçakaya M. Signal intensity informed multi-coil encoding operator for physics-guided deep learning reconstruction of highly accelerated myocardial perfusion CMR. Magn Reson Med. 2023 Jan;89(1):308–21.

119. Xue H, Tseng E, Knott KD, Kotecha T, Brown L, Plein S, et al. Automated detection of left ventricle in arterial input function images for inline perfusion mapping using deep learning: A study of 15,000 patients. Magn Reson Med. 2020 Nov;84(5):2788–800.

120. Xue H, Davies RH, Brown LAE, Knott KD, Kotecha T, Fontana M, et al. Automated Inline Analysis of Myocardial Perfusion MRI with Deep Learning. Radiol Artif Intell. 2020 Nov 1;2(6):e200009.

121. Jacobs M, Benovoy M, Chang LC, Corcoran D, Berry C, Arai AE, et al. Automated Segmental Analysis of Fully Quantitative Myocardial Blood Flow Maps by First-Pass Perfusion Cardiovascular Magnetic Resonance. IEEE Access. 2021;9:52796–811.

122. Yalcinkaya DM, Youssef K, Heydari B, Zamudio L, Dharmakumar R, Sharif B. Deep Learning-Based Segmentation and Uncertainty Assessment for Automated Analysis of Myocardial Perfusion MRI Datasets Using Patch-Level Training and Advanced Data Augmentation. In: 2021 43rd Annual International Conference of the IEEE Engineering in Medicine & Biology Society (EMBC) [Internet]. Mexico: IEEE; 2021 [cited 2023 Dec 22]. p. 4072–8. Available from: https://ieeexplore.ieee.org/document/9629581/

123. Kim YC, Kim K, Choe YH. Automatic calculation of myocardial perfusion reserve using deep learning with uncertainty quantification. Quant Imaging Med Surg. 2023;

124. García-Jara G, Jimenez-Molina A, Reyes E, Tapia-Rivas N, Ramos-Gómez C, De Grazia J, et al. Efficient and Motion Correction-Free Myocardial Perfusion Segmentation in Small MRI Data Using Deep Transfer Learning From Cine Images: A Promising Framework for Clinical Implementation. IEEE Access. 2023;11:103177–88.

125. Van Herten RLM, Chiribiri A, Breeuwer M, Veta M, Scannell CM. Physics-informed neural networks for myocardial perfusion MRI quantification. Med Image Anal. 2022 May;78:102399.

126. Alskaf E, Suinesiaputra A, Scannell CM, Razavi R, Ourselin S, Young A, et al. Hybrid Artificial Intelligence Outcome Prediction Using Feature Extraction from Stress Perfusion Cardiac Magnetic Resonance Images and Electronic Health Records. Eur Heart J. 2023 Nov 9;44(Supplement_2):ehad655.2934.

127. Demirel ÖB, Zhang C, Yaman B, Gulle M, Shenoy C, Leiner T, et al. High-fidelity Database-

free Deep Learning Reconstruction for Real-time Cine Cardiac MRI [Internet]. Bioengineering; 2023 Feb [cited 2023 Nov 23]. Available from: http://biorxiv.org/lookup/doi/10.1101/2023.02.13.528388

128. Morales MA, Izquierdo-Garcia D, Aganj I, Kalpathy-Cramer J, Rosen BR, Catana C. Implementation and Validation of a Three-dimensional Cardiac Motion Estimation Network. Radiol Artif Intell. 2019 Jul;1(4):e180080.

129. Martín-González E, Sevilla T, Revilla-Orodea A, Casaseca-de-la-Higuera P, Alberola-López C. Groupwise Non-Rigid Registration with Deep Learning: An Affordable Solution Applied to 2D Cardiac Cine MRI Reconstruction. Entropy. 2020 Jun 19;22(6):687.

130. Pan J, Rueckert D, Kustner T, Hammernik K. Efficient Image Registration Network for Non-Rigid Cardiac Motion Estimation.

131. Lyu Q, Shan H, Xie Y, Kwan AC, Otaki Y, Kuronuma K, et al. Cine Cardiac MRI Motion Artifact Reduction Using a Recurrent Neural Network. IEEE Trans Med Imaging. 2021 Aug;40(8):2170–81.

132. Qi H, Hajhosseiny R, Cruz G, Kuestner T, Kunze K, Neji R, et al. End-to-end deep learning nonrigid motion-corrected reconstruction for highly accelerated free-breathing coronary MRA. Magn Reson Med. 2021 Oct;86(4):1983–96.

133. Lv J, Yang M, Zhang J, Wang X. Respiratory motion correction for free-breathing 3D abdominal MRI using CNN-based image registration: a feasibility study. Br J Radiol. 2018 Jan 1;91(1083):20170788.

134. Tamada D, Kromrey ML, Ichikawa S, Onishi H, Motosugi U. Motion Artifact Reduction Using a Convolutional Neural Network for Dynamic Contrast Enhanced MR Imaging of the Liver. Magn Reson Med Sci. 2020;19(1):64–76.

135. Huang J, Guo J, Pedrosa I, Fei B. Deep learning-based deformable registration of dynamic contrast enhanced MR images of the kidney. In: Linte CA, Siewerdsen JH, editors. Medical Imaging 2022: Image-Guided Procedures, Robotic Interventions, and Modeling [Internet]. San Diego, United States: SPIE; 2022 [cited 2024 May 15]. p. 48. Available from: https://www.spiedigitallibrary.org/conference-proceedings-of-spie/12034/2611768/Deep-learning-based-deformable-registration-of-dynamic-contrast-enhanced-MR/10.1117/12.2611768.full

136. Gong Z, Liu Z, Guo W, Zhao D, He C, Liu Y, et al. Deep learning of deformable registration for breast DCE-MRI images. In: The Fourth International Symposium on Image Computing and Digital Medicine [Internet]. Shenyang China: ACM; 2020 [cited 2024 May 15]. p. 229–34. Available from: https://dl.acm.org/doi/10.1145/3451421.3451469

137. Aprea F, Marrone S, Sansone C. Neural Machine Registration for Motion Correction in Breast DCE-MRI. In: 2020 25th International Conference on Pattern Recognition (ICPR) [Internet]. Milan, Italy: IEEE; 2021 [cited 2024 May 15]. p. 4332–9. Available from: https://ieeexplore.ieee.org/document/9412116/

138. Balakrishnan G, Zhao A, Sabuncu MR, Guttag J, Dalca AV. VoxelMorph: A Learning Framework for Deformable Medical Image Registration. IEEE Trans Med Imaging. 2019 Aug;38(8):1788–800.

139. Qi H, Fuin N, Cruz G, Pan J, Kuestner T, Bustin A, et al. Non-Rigid Respiratory Motion Estimation of Whole-Heart Coronary MR Images Using Unsupervised Deep Learning. IEEE Trans Med Imaging. 2021 Jan;40(1):444–54.

140. Kustner T, Pan J, Qi H, Cruz G, Gilliam C, Blu T, et al. LAPNet: Non-Rigid Registration Derived in k-Space for Magnetic Resonance Imaging. IEEE Trans Med Imaging. 2021 Dec;40(12):3686–97.

141. Oksuz I, Clough J, Bustin A, Cruz G, Prieto C, Botnar R, et al. Cardiac MR Motion Artefact Correction from K-space Using Deep Learning-Based Reconstruction. In: Knoll F, Maier A, Rueckert D, editors. Machine Learning for Medical Image Reconstruction [Internet]. Cham: Springer International Publishing; 2018 [cited 2024 May 15]. p. 21–9. (Lecture Notes in Computer Science; vol. 11074). Available from: https://link.springer.com/10.1007/978-3-030-00129-2_3

142. Oksuz I, Clough JR, Ruijsink B, Anton EP, Bustin A, Cruz G, et al. Deep Learning-Based Detection and Correction of Cardiac MR Motion Artefacts During Reconstruction for High-Quality Segmentation. IEEE Trans Med Imaging. 2020 Dec;39(12):4001–10.

143. Yang J, Küstner T, Hu P, Liò P, Qi H. End-to-End Deep Learning of Non-rigid Groupwise Registration and Reconstruction of Dynamic MRI. Front Cardiovasc Med. 2022 Apr 28;9:880186.

144. Pan J, Hamdi M, Huang W, Hammernik K, Kuestner T, Rueckert D. Unrolled and rapid motion-compensated reconstruction for cardiac CINE MRI. Med Image Anal. 2024 Jan;91:103017.

145. Huang Q, Xian Y, Yang D, Qu H, Yi J, Wu P, et al. Dynamic MRI reconstruction with end-to-end motion-guided network. Med Image Anal. 2021 Feb;68:101901.

146. Yang J, Küstner T, Hu P, Liò P, Qi H. End-to-End Deep Learning of Non-rigid Groupwise Registration and Reconstruction of Dynamic MRI. Front Cardiovasc Med. 2022 Apr 28;9:880186.

147. Alskaf E, Suinesiaputra A, Scannell CM, Razavi R, Ourselin S, Young A, et al. Hybrid Artificial Intelligence Outcome Prediction Using Feature Extraction from Stress Perfusion Cardiac Magnetic Resonance Images and Electronic Health Records. Eur Heart J. 2023 Nov 9;44(Supplement_2):ehad655.2934.

148. Engblom H, Xue H, Akil S, Carlsson M, Hindorf C, Oddstig J, et al. Fully quantitative cardiovascular magnetic resonance myocardial perfusion ready for clinical use: a comparison between cardiovascular magnetic resonance imaging and positron emission tomography. J Cardiovasc Magn Reson. 2016 Dec;19(1):78.

149. American Heart Association Writing Group on Myocardial Segmentation and Registration for Cardiac Imaging:, Cerqueira MD, Weissman NJ, Dilsizian V, Jacobs AK, Kaul S, et al. Standardized Myocardial Segmentation and Nomenclature for Tomographic Imaging of the Heart: A Statement for Healthcare Professionals From the Cardiac Imaging Committee of the Council on Clinical Cardiology of the American Heart Association. Circulation. 2002 Jan 29;105(4):539–42.

150. Hsu LY, Jacobs M, Benovoy M, Ta AD, Conn HM, Winkler S, et al. Diagnostic Performance of Fully Automated Pixel-Wise Quantitative Myocardial Perfusion Imaging by Cardiovascular Magnetic Resonance. JACC Cardiovasc Imaging. 2018 May;11(5):697–707.

151. Benovoy M, Jacobs M, Cheriet F, Dahdah N, Arai AE, Hsu L. Robust universal nonrigid motion correction framework for first-pass cardiac MR perfusion imaging. J Magn Reson Imaging. 2017 Oct;46(4):1060–72.

152. Zhao S hai, Guo W feng, Yao Z feng, Yang S, Yun H, Chen Y yin, et al. Fully automated pixel-wise quantitative CMR-myocardial perfusion with CMR-coronary angiography to detect hemodynamically significant coronary artery disease. Eur Radiol. 2023 May 5;33(10):7238–49.

153. Xue H, Davies R, Hansen D, Tseng E, Fontana M, Moon JC, et al. Gadgetron Inline AI: Effective Model inference on MR scanner. In 2019.

154. Kotecha T, Chacko L, Chehab O, O'Reilly N, Martinez-Naharro A, Lazari J, et al. Assessment of Multivessel Coronary Artery Disease Using Cardiovascular Magnetic Resonance Pixelwise Quantitative Perfusion Mapping. JACC Cardiovasc Imaging. 2020 Dec;13(12):2546–57.

155. Scannell CM, Veta M, Villa ADM, Sammut EC, Lee J, Breeuwer M, et al. Deep-Learning-Based Preprocessing for Quantitative Myocardial Perfusion MRI. J Magn Reson Imaging. 2020 Jun;51(6):1689–96.

156. Brown LAE, Onciul SC, Broadbent DA, Johnson K, Fent GJ, Foley JRJ, et al. Fully automated, inline quantification of myocardial blood flow with cardiovascular magnetic resonance: repeatability of measurements in healthy subjects. J Cardiovasc Magn Reson. 2018 Feb;20(1):48.

157. Scannell CM, Crawley R, Alskaf E, Breeuwer M, Plein S, Chiribiri A. High-resolution quantification of stress perfusion defects by cardiac magnetic resonance. Eur Heart J - Imaging Methods Pract. 2024 Jan 16;2(1):qyae001.